\documentclass{article}[12pt]
\usepackage{jheppub}
\usepackage[utf8]{inputenc}
\usepackage{amsmath}
\usepackage{amssymb}
\usepackage{amsfonts}
\usepackage{amsthm}
\usepackage{xcolor}
\usepackage{colonequals}
\usepackage{rotating}
\usepackage{array}
\def\SO{\textrm{SO}}
\def\Spin{\textrm{Spin}}
\def\Pin{\textrm{Pin}}
\title{Conformal Group Theory of Tensor Structures}

\preprint{DESY 19-172}

\author{Ilija Buri\' c$^1$,}
\author{Volker Schomerus$^1$}
\author{and Mikhail Isachenkov$^2$}
\affiliation{$^1$DESY, Notkestra\ss e 85, D-22607 Hamburg, Germany}
\affiliation{$^2$IH\'ES, 35 Route de Chartres, 91440 Bures-sur-Yvette, France}

\abstract{The decomposition of correlation functions into conformal blocks is an
indispensable tool in conformal field theory. For spinning correlators, non-trivial
tensor structures are needed to mediate between the conformal blocks, which are
functions of cross ratios only, and the correlation functions that depend on
insertion points in the $d$-dimensional Euclidean space. Here we develop an
entirely group theoretic approach to tensor structures, based on the Cartan
decomposition of the conformal group. It provides us with a new universal
formula for tensor structures and thereby a systematic derivation of crossing
equations. Our approach applies to a `gauge' in which the conformal blocks are
wave functions of Calogero-Sutherland models rather than solutions of the more
standard Casimir equations. Through this ab initio construction of tensor
structures we complete the Calogero-Sutherland approach to conformal
correlators, at least for four-point functions of local operators in
non-supersymmetric models. An extension to defects and superconformal
symmetry is possible.}

\begin{document}

\maketitle

\vspace{1cm}

\addtolength{\baselineskip}{1mm}
\newpage
\section{Introduction}

Conformal partial wave expansions are an indispensable tool in conformal field theory
that allows to decompose correlation functions of local and non-local operators into a
basis of kinematically determined building blocks. Such decompositions were introduced
in the early days of conformal field theory, but the mathematical nature of the blocks
remained elusive for a long time. In fact, before the work of Dolan and Osborn
\cite{Dolan:2000ut,Dolan:2003hv}, the shadow formalism of Ferrara et al.\
\cite{Ferrara:1972uq} only provided certain integral expressions even for correlation
functions of scalar fields. Dolan and Osborn characterized the scalar blocks through
second order Casimir differential equations and thereby managed to uncover many of
their properties, including explicit formulas for scalar four-point blocks in even
dimensions. Boosted by the success of the numerical bootstrap \cite{Rattazzi:2008pe,
ElShowk:2012ht, El-Showk:2014dwa,Simmons-Duffin:2015qma,Kos:2016ysd} the construction
of conformal partial wave expansions was extended in many different directions.
Analyzing correlators of fields with spin, such as e.g.\ currents or the stress
tensor, is clearly one of the most pressing issues that was addressed early on in
\cite{Costa:2011dw,SimmonsDuffin:2012uy,Costa:2014rya,Costa:2016hju,
Costa:2016xah}.
\smallskip

While correlation functions depend on insertion points $x_i$, conformal blocks are
functions of cross ratios $z_i$ only. Hence, any conformal partial wave expansion
requires inserting a properly chosen factor that can turn functions of conformally
invariant cross ratios into functions of insertion points in such a way that conformal
Ward identities are satisfied. This factor $\Theta$, usually referred to as ``tensor
structures'', may be considered as a rectangular matrix whose entries are functions
of the insertion points. One index of the matrix $\Theta$ collects all the tensor
indices of the involved fields while the other runs through some basis in the space of
tensor structures. Conformal Ward identities do not determine the tensor structures
uniquely since they are left invariant if we multiply these by a
matrix of functions of the cross ratios. This inherent gauge freedom is usually
fixed by making some more or less convenient or simple choices before working
out the Casimir differential equations and constructing the associated conformal
blocks, see e.g.\ \cite{Iliesiu:2015akf,Echeverri:2016dun} for some examples.
\smallskip

In \cite{Isachenkov:2016gim} the authors noticed that the Casimir equations for
scalar four-point functions derived by Dolan and Osborn could be (gauge) transformed into
stationary Schroedinger equations for integrable two-particle Hamiltonians of
Calogero-Sutherland type. The latter have been studied extensively in the mathematical
literature, starting with the work of Heckman and Opdam \cite{Heckman-Opdam}. The
connection uncovered in \cite{Isachenkov:2016gim} implies that conformal blocks are
certain multivariable hypergeometric functions with properties that follow from
integrability of the quantum mechanical system and are very similar to those of
ordinary hypergeometrics in one variable. Later, it was understood that the relation
between conformal blocks and Calogero-Sutherland models is entirely universal. More
concretely, for all spinning correlators there exists some gauge in which the associated
Casimir equations possess Calogero-Sutherland form, see \cite{Schomerus:2016epl,
Schomerus:2017eny}. In contrast to the standard approach we described above, the
associated Casimir equations could be constructed without ever fixing particular
tensor structures $\Theta$. This underlines the canonical nature of the
Calogero-Sutherland gauge in comparison to all other choices and it has made it
easy to detect relations between seemingly different setups, e.g.\ between Casimir
equations for four-point functions and defect correlators \cite{Isachenkov:2018pef},
and to apply standard solution generating techniques \cite{Isachenkov:2017qgn,
Isachenkov:2019xxx} to study new types of blocks.
\medskip

In all cases in which Casimir equations for spinning conformal blocks had been
known, they were shown by explicit computations to be equivalent to eigenvalue
equations for spinning Calogero-Sutherland Hamiltonians. The equivalence is
mediated by a specific gauge transformation that relates the two sets of
Casimir equations. Once the gauge transformation is known it can be used
to establish a direct relation between blocks in Calogero-Sutherland gauge
and correlation functions since tensor structures for many kinds of spinning
blocks had been worked out explicitly. On the other hand, it is certainly
not satisfactory that so far the explicit relation between correlation
functions and Calogero-Sutherland wave functions was composed from previously
constructed tensor structures and a gauge transformation between usual Casimir
equations and Calogero-Sutherland eigenvalue equations. More importantly, in
the context of superblocks, there exists a large class of cases now in which
the Calogero-Sutherland eigenvalue equations are known while tensor structures
and the associated Casimir equations have not been constructed. In those cases
a direct construction of tensor structures in Calogero-Sutherland gauge is
indispensable. Let us note that also the conformal bootstrap programme cannot
do without this. Even though the crossing symmetry constraints are usually written
for functions of cross ratios only, knowledge of $s$- and $t$-channel blocks
is not sufficient. To set up the relevant equations one also has to compute
the ratio of $s$- and $t$-channel tensor structures. The resulting crossing
matrix is a non-trivial function of the cross ratios that must be taken into
account when solving for the dynamical content of the theory.
\smallskip

All this provides strong motivation for the present work in which we will construct
a direct map between correlation functions and Calogero-Sutherland wave functions.
We have explained above that Calogero-Sutherland eigenvalue equations arise
when we write Casimir equations for conformal blocks in a distinguished
gauge, one that is deeply routed in the group theory of the conformal
group. In this context, we may rephrase the main result of this work as
a direct construction of tensor structures in the very same canonical
Calogero-Sutherland gauge. As we shall see below, tensor structures in
the Calogero-Sutherland gauge arise from a change of coordinates on the
conformal group. This group theoretic understanding will allow us to
provide universal formulas for tensor structures. More concretely we will
prove that any spinning four-point correlation function $G$ possesses a
canonical factorization of the form
\begin{equation} \label{eq:result}
G(\pi_i;x_i) = \tilde \Omega(\Delta_i;x_i) \Theta(\lambda_i;x_i)
\psi(\pi_i;u_1,u_2)\ .
\end{equation}
Here $\pi_i = (\Delta_i,\lambda_i)$ denote the conformal weights $\Delta_i$ and
spins $\lambda_i$ of the four fields.
The first two factors are split off so that $\psi$ depends on the insertion
points only through the cross ratios $z=z_1$ and $\bar z = z_2$. The arguments
$u_i$ we display here are related to usual cross ratios $z_i$ as $z_i = -
\sinh^{-2} u_i/2$.  Our concrete choice for $\tilde \Omega$ and $\Theta$, see
below, ensures that $\psi$ can be expanded in eigenfunctions of some matrix
valued Calogero-Sutherland Hamiltonian with constant coefficients. The
eigenvalue equation for the Calogero-Sutherland Hamiltonian takes the place
of the conventional Casimir equations for conformal blocks $g$.

The prefactor $\tilde \Omega \Theta$ - which ensures the relation with
Calogero-Sutherland models - is composed from two factors, one scalar factor
$\tilde \Omega$ that depends only on the conformal weights $\Delta_i$ and
another matrix valued factor that is sensitive to the spins $\lambda_i$.
Put differently, while $\tilde \Omega$ is associated with the transformation
under dilations $d \in D = \SO(1,1)$, the tensor structures $\Theta$
mediate the transformation behavior with respect to the action of the (double covering of the)
rotation group $R = \Spin(d)$. Our choice of the function $\tilde \Omega$ is
easy to spell out. In the $s$-channel, for example, it takes the form
\begin{equation}
\tilde \Omega_s(x_i) = \frac{\frac12 (-1)^\frac{d-2}{2}}{x_{12}^{\Delta_1+\Delta_2}
x_{34}^{\Delta_3+\Delta_4}}\left(\frac{x_{14}}{x_{24}}\right)^{2a} \left(\frac{x_{14}}
{x_{13}}\right)^{2b}\frac{(z_1z_2)^\frac{d-1}{2}|z_1-z_2|^{-\frac{d-2}{2}} }
{[(1-z_1)(1-z_2)]^{\frac{a}{2}+\frac{b}{2}+\frac{1}{4}}}\ .
\end{equation}
This differs from the function $\Omega$ that was introduced in
\cite{Dolan:2003hv} by some specific function of the cross ratios
which was first constructed in \cite{Isachenkov:2016gim}. The most
novel part of our formula \eqref{eq:result} is the construction of
the tensor structures. In order to obtain the matrix of tensor
structures $\Theta = \Theta_s$ in the $s$-channel, for example, we
consider the following family of elements of the conformal group,
\begin{equation} \label{eq:result2}
g(x_i)=e^{\frac{x^\mu_{21}}{x_{21}^2}K_\mu}e^{x^\mu_{13} P_\mu}
e^{\frac{x^\mu_{34}}{x_{34}^2}K_\mu} = k_l(x_i)\,
e^{\frac{u_1+u_2}{4}(P_1+K_1) - i\frac{u_1-u_2}{4}(P_2 - K_2)} \,
k_r(x_i)\ .
\end{equation}
It turns out that almost all elements of the conformal group can be
written as a product of an element $k_l \in K = \SO(1,1) \times
\Spin(d)$, an element $a$ of a particular 2-dimensional abelian subgroup
and another element $k_r \in K$, see the right hand side of the previous
equation. This decomposition defines two functions $k_l(x_i)$ and
$k_r(x_i)$ of the insertion points that take values in the subgroup $K$.
When these are represented in some finite dimensional representations
of the rotation group, one that is determined by the set of spins
$\lambda_i$, we obtain the tensor structures $\Theta_s$. A concrete
formula will be stated in Section 3.
\medskip

Let us now describe the plan of this paper in more detail. In the next
section we shall explain how to rewrite conformal correlation functions
as functions on the conformal group. This step may be compared to writing
correlation functions in embedding space. In particular, the space of
functions on the conformal group also admits a linear action of conformal
symmetry. Once we have lifted the correlator to the conformal group, we
can work out the Casimir equations through an appropriate restriction of
the Laplace operator on the group. This employs a very particular
set of coordinates which can be considered as higher dimensional
generalizations of the usual Euler angles for $\SO(3)$ (or rather
$\SO(1,2)$). Since the new coordinates arise from a KAK, or Cartan,
decomposition of the conformal group we shall sometimes refer to them 
as Cartan coordinates. These Cartan coordinates
make it straightforward to restrict the Laplace operator to those functions
on the group that are relevant for conformal correlation functions.
After this restriction we obtain the Calogero-Sutherland Hamiltonians we
mentioned before. In this context, the tensor structures simply arise
from a change of coordinates on the conformal group. Initially we
parametrize group elements through the insertion points of our local
fields, as stated in eq.\ \eqref{eq:result2}. The transformation to
Cartan coordinates is a group theoretic problem that, once it
is solved, provides universal formulas for the tensor structures, for
any given assignment of spins. The general theory will be explained in
Section 3, before we describe the explicit solution and computation of
tensor structures for the conformal groups in $d=3$ and $d=4$ dimensions
in Section 4. In Section 5 we evaluate the tensor structures for some special
choices of spins to compare with previous expressions in the literature, most
notably in \cite{Iliesiu:2015akf} and \cite{Echeverri:2016dun}. Our work
concludes with an outlook. In particular we compute the crossing matrix for
the simplest spinning seed blocks in $d=4$ dimensions and observe that is
takes a very simple form. In addition we comment  briefly on possible
extensions to superblocks  and defects.

\section{Correlators as Functions on the Conformal Group}

In this section, our main goal is to reinterpret four-point correlation functions
of local primary fields in a conformal field theory in terms of functions on the
conformal group which obey certain covariance properties. In the first
subsection we collect some background material on primary fields, correlation
functions and Ward identities. The discussion includes a review of conformal and
Weyl inversions. The main result of the section is stated at the beginning of the
second subsection, along with a sketch of the derivation and some additional
comments on the relation with embedding space constructions. A full
proof of our result can be found in Appendices A and B.

\subsection{Background, conformal and Weyl inversions}

A local primary field $\Phi_{(\Delta,\lambda)}$ in conformal field theory is
characterized by its weight $\Delta$ and its spin $\lambda$. In representation
theoretic terms, the weight and spin correspond to a finite dimensional representation
$\pi=(\Delta,\lambda)$ of the product group $K = \SO(1,1) \times \Spin(d) \subset G =
\Spin(1,d+1)$. By definition, the transformation behavior of such a conformal primary
field with respect to infinitesimal conformal transformations is given by
\begin{eqnarray}
[D,\Phi_\pi(x)] & = & \left(x^{\nu} \partial_{\nu}+\Delta\right) \Phi_\pi (x) \\[2mm]
[L_{\mu\nu},\Phi_\pi(x)] & = & \left(x_{\nu} \partial_{\mu}- x_{\mu}
\partial_{\nu}+\Sigma{}^\lambda_{\mu \nu} \right) \Phi_\pi(x) \label{eq:primary}\\[2mm]
[K_{\mu}, \Phi_\pi(x)]&= & \left( x^{2} \partial_{\mu} - 2 x_{\mu} x^{\nu}
\partial_{\nu}-2 x_{\mu} \Delta + 2 x^{\nu} \Sigma{}^\lambda_{\mu \nu}\right) \Phi (x)
\end{eqnarray}
where $\Sigma{}^\lambda_{\mu\nu}= \pi (L_{\mu\nu}) $ denote the representation matrices
of the generators $L_{\mu\nu}$ of rotations in the representation $\lambda$ of $\Spin(d)$.
The generators of translations act simply as derivatives.

It is important for us to understand the action of the conformal inversion in some detail.
The conformal inversion $I: \mathbb{S}^d \xrightarrow{}\mathbb{S}^d$
acts on the coordinates of the conformally compactified Euclidean space $\mathbb{S}^d$
as
\begin{equation}
    I(x^\mu) = \frac{x^\mu}{x^2} \ . \nonumber
\end{equation}
In order to determine the transformation law of primary fields under the conformal inversion,
we are instructed to compute the differential (Jacobian) of $I$, which takes the form
\begin{equation}
    dI^\mu_\nu(x) = \partial_\nu \frac{x^\mu}{x^2}  = \frac{1}{x^2}(\delta^\mu_\nu -
    2\hat x^\mu \hat x_\nu) =  \frac{1}{x^2} s_x \ . \nonumber
\end{equation}
Thus the Jacobian $dI$ decomposes as the product of a conformal factor $x^{-2}$ and a
reflection $s_x$ in the hyperplane orthogonal to $x$. The inversion $I$ can be thought
of as an element of $Pin(1,d+1)$ with $\textit{det}(I)=-1$, i.e.\ it does not belong to
the identity component of $Pin(1,d+1)$. Explicitly, the inversion $I$ takes the form
$I=\textit{diag}(1,-1,1, \dots, 1)$. In a parity-invariant conformal field theory, a
primary field which transforms in the representation $\pi$ of the orthogonal group
$O(d)$ is acted on by $I$ as
\begin{equation}
  \Phi^I(x) = I \Phi(x) I^{-1} =
    \Big(\frac{1}{x^2}\Big)^{-\Delta} \pi(s_x) \Phi(I(x)). \nonumber
\end{equation}
Since in a parity-invariant theory fields only transform in those representations $\pi$ of the
rotation group $\Spin(d)$ which extend to $\Pin(d)$, the representation
matrix $\pi(s_x)$ is well-defined. To describe the transformation behavior in a general
conformal field theory, we introduce the reflections $s_{e_i}$ along the hyperplane
orthogonal to $e_i$
\begin{equation}
    s_{e_i}: (x_1,\dots,x_i,\dots,x_n)\mapsto(x_1,\dots,-x_i,\dots,x_n). \nonumber
\end{equation}
With the help of these reflections, we can now define new maps $I_i=Is_{e_i} = s_{e_i}I$
which are in the identity component $\Spin(1,d+1)$ of the conformal group. Their
differentials are given by
\begin{equation}
    dI_i(x) = \frac{1}{x^2}s_{e_i} s_{x} \ .
\end{equation}
We will denote the group element $I_d$ by $w$ and call it the Weyl inversion. Under
the action of $w$, a primary field transforms as
\begin{equation} \label{eq:Phiinv}
\Phi^{w}(x) =  \Big(\frac{1}{x^2}\Big)^{-\Delta} \pi(s_{e_d} s_x) \Phi
(w(x))\ .
\end{equation}
We note that the product $s_{e_d}s_{x}$ of two reflections is a proper rotation and hence its
representation matrix $\pi(s_x s_{e_d})$ is well defined for all representations $\pi$ of
the spin group. For later use let us note that the Weyl inversion $w$ can be used to
construct a new representation $\pi' = \pi^w$ of the subgroup $K$ from $\pi$ through
$$ \pi' (k) :=  \pi (w k w^{-1}) \  $$
for all  $k\in K$. By construction $w k w^{-1}\in K$ and hence these elements can be
evaluated in $\pi$ so that the right hand side is well defined. In general, the
representations $\pi$ and $\pi'$ are inequivalent. We also observe that
\begin{equation} \label{eq:piprimeD}
 \pi' (D) = -\pi(D) \ .
\end{equation}
This concludes our short detour on conformal inversions and their action on primary
fields in a general conformal field theory.
\smallskip

Let us now choose four representations $\pi_i=(\Delta_i,\lambda_i),\ i=1, \dots,4$, one
for each primary field we want to insert, and consider the correlation function
\begin{equation}\label{eq:G}
G(\pi_i,x_i) = \langle \Phi_{\pi_1}(x_1) \dots \Phi_{\pi_4} (x_4) \rangle.
\end{equation}
We shall think of this as a function from
the set of insertion points $x_i$ to the space
\begin{equation} \label{eq:W}
 W_P = \bigotimes_{i=1}^4 V_i\
\end{equation}
of polarisations. Here, $V_i$ denotes the finite dimensional carrier space of the
representation $\pi_i$ of $K$. Conformal invariance implies that the four-point
functions $G$ satisfy a set of differential Ward identities which take the form
\begin{equation}
\langle [X,\Phi_{\pi_1}(x_1) \dots \Phi_{\pi_4}(x_4)] \rangle = 0
\end{equation}
for all generators $X$ of the conformal algebra. The commutator of $X$ with the
individual primary fields can be evaluated using the familiar commutation
relations \eqref{eq:primary}. In the case of dilations, i.e.\ for $X=D$, for
example, the Ward identity reads
\begin{equation} \label{eq:WardD}
\sum_{i=1}^4 x_i^\nu \partial^{(i)}_\nu G(x_i) = - \sum_{i=1}^4
\Delta_i G(x_i)\
\end{equation}
where $\partial^{(i)}$ denotes the derivative with respect to the insertion point
$x_i$. For the other generators, the Ward identities take a similar form.

\subsection{Correlators from covariant functions on the group}

Our goal in this subsection is to show that, for any given choice of a channel, the
solutions of the Ward identities are in 1-1 correspondence with certain covariant
functions on the conformal group. The representation theoretic data we have introduced,
i.e.\ the four representations $\pi_i = (\Delta_i,\lambda_i)$ of the subgroup $K$,
determine a subspace $\Gamma^{\pi_i}_{\sigma}$ of functions on the conformal group
that take values in the space $W_P$ of polarisations. In order to characterize the
relevant space of functions we introduce the representation $\hat \pi_\sigma$
as\footnote{Here and in the following the products such as $K\times K$ of subgroups
  of $G$ are understood as products over $\mathbb{Z}_2$, where $\mathbb{Z}_2$ is a
  kernel of the projection $\Spin(1, d+1)\twoheadrightarrow \SO_e(1, d+1)$.}
\begin{eqnarray}
\hat \pi_\sigma (k_l,k_r) & = & \pi_{\sigma(1)\sigma(2)}(k_l)
\otimes \pi_{\sigma(3)\sigma(4)}(k^{-1}_r)\ \quad \textit{for} \quad (k_l,k_r) \in
  K \times K \label{eq:hatpi} \\[2mm]
\textit{where} & &  \pi_{ij}(k) =   \pi_i(k) \otimes \pi'_j(k) \quad \textit{for}
\quad k \in K
\end{eqnarray}
and $\sigma \in S_4$ is a permutation of four points. With this notation we can now
define the linear subspace $\Gamma^{\pi_i}_{\sigma}$ to consist of all those $W_P$-valued
functions on the conformal group which satisfy the covariance laws
\begin{equation} \label{covariance}
F_\sigma\left(k_l g k_r\right) = \hat \pi_\sigma(k_l,k_r) F_\sigma(g) \ .
\end{equation}
Within this subspace, whose definition depends on the choice of a channel
$\sigma \in S_4$, we can now find a unique representative of the correlation
function $G(x_i)$. More precisely, given any solution $G(x_i)$ of the Ward identities,
there exists a unique function $F_\sigma\in\Gamma^{\pi_i}_\sigma$ such that
\begin{equation} \label{main}
G(x_i) = \frac{1}{x_{\sigma(1)\sigma(2)}^{2\Delta_{\sigma(2)}}
x_{\sigma(3)\sigma(4)}^{2\Delta_{\sigma(4)}}} \ \hat \rho_\sigma (x_i)\
F_\sigma\left( e^{Ix_{\sigma(2)\sigma(1)}\cdot K}e^{x_{\sigma(1)\sigma(3)}\cdot P}
e^{Ix_{\sigma(3)\sigma(4)}\cdot K}\right)
\end{equation}
where $\sigma \in S_4$ and we introduced a rotation matrix $\hat \rho_\sigma$
that is defined as
\begin{eqnarray}
\hat \rho_\sigma (x_i) & = & \rho_{\sigma(1)\sigma(2)}(x_{\sigma(1)},x_{\sigma(2)})
\otimes \rho_{\sigma(3)\sigma(4)}(x_{\sigma(3)},x_{\sigma(4)})\
\label{eq:hatrho} \\[2mm]
\textit{where} & &  \rho_{ij}(x_i,x_j) =   \pi_i(e) \otimes \pi_j(s_{x_{ij}}s_{e_d}) .
\end{eqnarray}
The permutation group $S_4$ consists of $|S_4| = 4! = 24$ elements. The permutations
$\sigma = (1,2), \sigma = (3,4)$ and $\sigma =(1,3)(2,4)$ generate a subgroup of order
eight whose action on the individual objects on the right hand side of eq.\ \eqref{main}
is almost trivial. The quotient of $S_4$ with respect to this subgroup consists of three
classes. We shall denote these by $s,t$ and $u$ with $\sigma_s = \textit{id}$, $\sigma_t
= (2,4)$ and $\sigma_u = (2,3)$. In proving formula \eqref{main} we will restrict to the
$s$-channel. All other cases can be dealt with similarly. For $\sigma = \sigma_s
= \textit{id}$ our equation \eqref{main} reads
\begin{equation} \label{maine}
G(x_i) = \frac{1}{x_{12}^{2\Delta_2}x_{34}^{2\Delta_4}}(1\otimes\pi_2(s_{x_{12}}s_{e_d})
\otimes1\otimes\pi_4(s_{x_{34}}s_{e_d}))
F_s\left( e^{Ix_{21}\cdot K}e^{x_{13}\cdot P}e^{Ix_{34}\cdot K}\right).
\end{equation}
We establish our claim in two steps. First we need to show that the function on the
right hand side of eq.\ \eqref{maine} does indeed satisfy all Ward identities. Once we
are done with this step it remains to convince ourselves that all solutions of the
Ward identities admit a representation of the form \eqref{maine}. Here we want to carry
out the first step in the case of translations, dilations and rotations, postponing the
full proof to Appendix A. Translation symmetry is manifest since the right hand side
only depends on differences of insertion points. In order to derive the Ward identity
\eqref{eq:WardD} for dilations one makes use of the fact that
$$
\sum_{i=1}^4 x_i^\nu\partial^{(i)}_\nu e^{x_{13}\cdot P} = [D,e^{x_{13}\cdot P}]
\quad , \quad
\sum_{i=1}^4 x_i^\nu\partial^{(i)}_\nu e^{Ix_{21}\cdot K} = [D,e^{Ix_{21}\cdot K}]
\ .
$$
Hence, the action of the derivatives on the argument of the function $F_s$ may be
written as a commutator with the dilation generator. Both terms of the commutator
can be evaluated using an infinitesimal version of the covariance law
\eqref{covariance} along with the property \eqref{eq:piprimeD} so that we
obtain
$$
\sum_{i=1}^4 x_i^\nu \partial^{(i)}_\nu F_s\left( e^{Ix_{21}\cdot K}e^{x_{13}\cdot P}
e^{Ix_{34}\cdot K}\right) = (-\Delta_1 + \Delta_2 - \Delta_3 + \Delta_4)
F_s\left( e^{Ix_{21}\cdot K}e^{x_{13}\cdot P} e^{Ix_{34}\cdot K}\right)\
$$
where we also inserted $\pi_i(D) = - \Delta_i$. With the additional contributions
from the coefficient of $F_s$ in eq.\ \eqref{maine} we indeed find the desired
Ward identity \eqref{eq:WardD}. The Ward identities for rotations can be verified in
the same way. Once again we act with the infinitesimal generators of rotations on the
right hand side of equation \eqref{maine}. There are three terms that contribute,
namely the reflections $s_{x_{12}}$ and $s_{x_{34}}$ in the prefactor and the argument
of $F_s$. As in our discussion of dilations, the action on the argument of $F_s$ can be
represented through a commutator with the generator $L_{\mu\nu}$. The latter may be pulled
out of the argument using the covariance law \eqref{covariance}. After this, two of the
representation matrices, namely $\Sigma^2_{\mu\nu}$ and $\Sigma^4_{\mu\nu}$ are conjugated
with the element $w$. Since the conformal inversion commutes with rotations, we can
replace the conjugation with $w$ by a conjugation with the reflection $s_{e_d}$. This
conjugation with $s_{e_d}$ is removed when we commute these matrices past $\pi_2(s_{e_d})$ and
$\pi_4(s_{e_d})$. Commuting the representation matrices past the $x$-dependent reflections
yields two terms which cancel those that were obtained by acting the infinitesimal
generators on the prefactor. Putting all this together one can show that the right hand
side of eq.\ \eqref{maine} behaves in the same way under rotations as the correlator $G$.
More details can be found in Appendix A which contains a full proof of the conformal Ward
identities for the right hand side of eq.\ \eqref{maine}. The second step in our proof,
namely to establish existence of an appropriate function $F_s$ for any choice of the
correlation function $G$, is carried out in Appendix B.
\medskip

Let us conclude this section with a few more comments on eq.\ \eqref{maine}. In order
to appreciate the details a bit better, it is useful to think of the argument of the
function $F_s$ as a product $g = g_l g_r^{-1}$ of two elements
$$ g_l = e^{Ix_{21}\cdot K}e^{x_{1}\cdot P}\quad , \quad
g_r = e^{Ix_{43}\cdot K} e^{x_{3}\cdot P}\ . $$
Both $g_l$ and $g_r$ are elements of the conformal group $G= \Spin(1,d+1)$, though of a
quite special form. Since they are generated from translations and special conformal
transformations only, they can be thought of as parametrizing the coset space $K
\backslash G$ in which rotations and dilations are trivial. Thereby, the coset space $K
\backslash G$ may be considered as the configuration space of a pair of points in the
conformal compactification of the Euclidean space. There exists an intimate relation
with embedding space constructions. Recall that a point in Euclidean space is
represented by a light-like line in embedding space $\mathbb{R}^{1,d+1}$. Hence, any
two points in Euclidean space determine a pair of light-like lines and hence a
2-dimensional plane $P\subset \mathbb{R}^{1,d+1}$ in embedding space. For the first
two points $x_1$ and $x_2$ the corresponding plane $P_{x_1,x_2}$ is actually encoded
in the element $g_l$. More concretely, if we represent $g_l$ in the $d+2$-dimensional
(fundamental) representation of the conformal group $G$, the first two
rows of this matrix -- which consist of a light-like and a time-like vector in a $d+2$-%
dimensional Minkowski space --  span the plane $P_{x_1,x_2}$ in embedding space. The
remaining $d$ rows, span the orthogonal complement $P^\perp_{x_1,x_2}$ in embedding
space. A change of basis in 2-dimensional subspace $P$ and its orthogonal complement
$P^\perp$ is mediated by left multiplication with elements $k_l \in K = \SO(1,1)
\times \Spin(d)$. The action of conformal transformations on the pair of points $x_1$
and $x_2$, on the other hand, is given by the right action of group $G$ on the coset $K
\backslash G$. All this is true for the other two insertion points $x_3,x_4$ and the
group element $g_r$ we construct from these two points. The argument $g_l g_r^{-1}$
of our function $F_s$ in eq.\ \eqref{maine} therefore admits a left action of $K_l
\cong K$ and a right action of $K_r \cong K$. The right action of the conformal
group on $g_l \in K \backslash G$ and the left action on $g_r^{-1} \in G/K$ are
inverse to each other so that the element $g_l g_r^{-1} \in K \backslash G / K$ is
conformally invariant. Note that the double coset $K \backslash G /K$ is 2-dimensional,
in agreement with the fact that one can form two independent conformal invariants
from four points in $\mathbb{R}^d$. Since our functions $F_s$ satisfy the covariance
law \eqref{covariance} under the action of $K_l \times K_r$, the section $F_s(g_l
g_r^{-1})$ is entirely determined by the values it takes on the two-dimensional
space of conformal invariants, just as it is the case for conformal correlators
$G$. These comments may help to appreciate the content of the main result
\eqref{maine} in this section.

\section{Cartan Coordinates and Tensor Structures}

As it stands the formula (\ref{maine}) may not seem very useful. All we have done
was to express a function $G$ satisfying a set of covariance laws, the conformal
Ward identities, in terms of a $K_l \times K_r$ covariant function $F$ on
the conformal group. In this section, we will use the Cartan decomposition of the
conformal group to solve the $K_l \times K_r$ covariance laws. Geometric properties
of this decomposition ensure that the Casimir equations take the form of a matrix
Calogero-Sutherland system.

As described in the introduction our analysis will allow us to decompose the
correlation function $G$ into a product of some scalar function $\tilde \Omega$,
the tensor structures $\Theta$ and some function $\psi$ of two variables $u_1$ and
$u_2$. The choice of these coordinates and the form of the associated Casimir
equations will be reviewed briefly in the first part before we turn to the other
two factors in the second subsection. We will determine $\tilde \Omega$ completely
and from first principles and explain how the tensor structures $\Theta$ are
computed, leaving fully explicit formulas to the subsequent sections.

\subsection{Cartan coordinates and Calogero-Sutherland models}

In order to evaluate the right hand side of equation \eqref{maine} further it is
advantageous to change coordinates and work with a coordinate system that arises
from the KAK or Cartan decomposition of the conformal group, as explained in
\cite{Schomerus:2016epl,Schomerus:2017eny}. The Cartan coordinates generalize
the usual Euler angles for $\SO(3)$ to the $d$-dimensional conformal group in a
way that is ideally adapted to the covariance law \eqref{covariance} we
discussed in the previous section.

The Cartan decomposition is based on the decomposition of the Lie algebra
\begin{equation}
\mathfrak{g} = \mathfrak{k}\oplus \mathfrak{p}, \nonumber
\end{equation}
where $\mathfrak{k}=\text{Lie}(K)$ and $\mathfrak{p}$ is its orthogonal complement with respect to the Killing form.
The latter is spanned by translations $P_\mu$ and special conformal transformations
$K_\mu, \mu = 1, \dots, d$. Let us select a 2-dimensional abelian subalgebra
$\mathfrak{a}$ of $\mathfrak{p}$ that is spanned by $A_+ = (P_1+K_1)/2$ and $A_- =
(P_2-K_2)/2$. Through exponentiation we pass to the abelian subgroup $A\subset
\textrm{Spin}(1,d+1)$ that consists of elements of the form
\begin{equation}
\label{adef}
a(u_1,u_2) =  e^{\frac{u_1+u_2}{2} A_+ -i \frac{u_1-u_2}{2} A_-} =
 e^{\frac{u_1+u_2}{4}(P_1+K_1) -i \frac{u_1-u_2}{4}(P_2 - K_2)}\ .
\end{equation}
Almost all elements of the conformal group can be factorized as
\begin{equation}\label{eq:Cartan}
  g = k_l\,  a(u_1,u_2)\,  k_r \
\end{equation}
with $k_l, k_r \in K$. This factorization is not unique, however, since elements
of $A$ commute with $b\in B =\Spin(d-2) \subset \Spin(d)$. Consequently, the group element
$g$ is invariant under the action $(k_l,k_r) \mapsto (k_l b , b^{-1} k_r)$ of $\Spin(d)$
on $K \times K$. We can fix this gauge freedom by choosing a section $k_l \in K/B$.
\medskip

Let us now insert the Cartan decomposition \eqref{eq:Cartan} into the right hand side
of our representation \eqref{main} of the correlation function $G$. Using the covariance
covariance law \eqref{covariance} we see that
\begin{equation} \label{eq:Fpif}
F_\sigma (g) = \hat \pi_\sigma(k_l,k_r) f_\sigma(u_1,u_2)
\end{equation}
where $f(u_1,u_2)$ denotes the restriction of $F_\sigma$ from the conformal
group $G$ to the elements \eqref{adef} of the torus $A \subset G$. The map $\hat
\pi_\sigma$ was defined in eq.\ \eqref{eq:hatpi}. The factorization \eqref{eq:Fpif}
of the function $F_\sigma$ into a matrix $\hat \pi$ on $K_l \times K_r$ and a
function $f_\sigma$ of two variables is key to our discussion in this section.
Most of this work will focus on the matrix $\hat \pi$. The function $f_\sigma$,
on the other hand, was analysed in much detail in \cite{Schomerus:2016epl} and
\cite{Schomerus:2017eny}. We will use the rest of this subsection to review the
results and in particular precise relation of $f$ with Calogero-Sutherland wave
functions which we mentioned above.
\smallskip

In order to explain the mathematical nature of $f$ we start from the Laplacian
$\Delta^{(2)}_G$ on the conformal group. From the start, $\Delta^{(2)}_G$ acts
as a second order differential operator on functions $F$ which are defined
for any $g \in G$. By construction, however, it is invariant under both left and
right actions of the conformal group on itself and hence under the action of $K_l
\cong K$ and $K_r \cong K$, in particular. Consequently, the Laplacian descends to
a second order differential operator on the space of covariant functions
$\Gamma^{\pi_i}_\sigma$. Due to the Cartan decomposition, all functions in this space are
determined by their restrictions to the subgroup $A$. Since the decomposition suffers
from the ambiguity of shifting elements $b \in B = \Spin(d-2)$ between $k_l$ and $k_r$,
the function $f$ can only take values in a subspace $W_P^B \subset W_P$ of $B$
invariants in the space $W_P$ of all polarizations. The action of the Laplacian
on the restriction $f$ may be worked out explicitly and contains both second and
first order derivatives with respect to the two coordinates $u_1$ and $u_2$. Thanks
to the fact that the Cartan decomposition is hyperpolar \cite{Schomerus:2017eny},
one can eliminate the first order terms through a simple transformation of the
form\footnote{The numerical prefactor here is fixed from matching normalization
  of Dolan-Osborn for the case of scalar blocks \cite{Dolan:2011dv} with the
  standard normalization of the Harish-Chandra wavefunctions in Calogero-Sutherland
  theory.}
\begin{eqnarray}
&f_\sigma(u_1,u_2) = \omega^{-1/2}(u_1,u_2) \psi_\sigma(u_1,u_2)
\label{eq:fomegapsi} \\[4mm]
\textit{where}\quad & \omega(u_1,u_2) = 4(-1)^{2-d}
(\sinh\frac{u_1}{2} \sinh\frac{u_2}{2})^{2d-2}\coth\frac{u_1}{2}
\coth\frac{u_2}{2} |\sinh^{-2}\frac{u_1}{2}-\sinh^{-2}\frac{u_2}{2}|^{d-2}. \nonumber
\end{eqnarray}
Putting eqs.\ \eqref{eq:Fpif} and \eqref{eq:fomegapsi} together we
can write
$$ F_\sigma(g) = \Xi^{\hat \pi}_\sigma(g) \psi_\sigma (u_1,u_2) \quad
\textit{ where } \quad \Xi^{\hat \pi}_\sigma(g) =
\omega(u_1,u_2)^{-1/2} \hat \pi_\sigma(k_l,k_r):
W_P^B \mapsto W_P \ .
$$
The product $\omega \hat \pi$ certainly acts on the entire space $W_P$ of
polarizations. On the other hand, $f_\sigma$ only takes values in the $B=\Spin(d-2)$
invariant subspace of tensor structures. For this reason we want to think of $\Xi$
as a map from tensor structures to polarizations. Our discussion about the relation
between the Laplacian $\Delta^{(2)}_G$ on the conformal group $G$ and
Calogero-Sutherland Hamiltians $H^{\textit{CS}}$ can now be written as
\begin{equation}\label{eq:LapintHam}
\Delta^{(2)}_G \, \Xi^{\hat\pi}_\sigma(g) = \Xi^{\hat \pi}_\sigma(g)
\left[ \frac12 H^{\textit{CS}}_{\hat \pi,\sigma} + \frac{1}{8} (d^2-2d+1)
\right].
\end{equation}
Here $\Delta_G$ is a second order differential operator in all $(d+2)(d+1)/2$
coordinates on the $d$-dimensional conformal group $G = \Spin(1,d+1)$. The whole
equation is valid when applied to functions of two coordinates $u_1$ and $u_2$.
The Hamiltonian $H^\textit{CS}$ takes the form
$$ H^\textit{CS}_{\hat\pi,\sigma} = - \frac{\partial^2}{\partial u^2_1} -
\frac{\partial^2}{\partial u^2_2} + V^\textit{CS}_{\hat \pi,\sigma}(u_1,u_2)
$$
where $V$ is a matrix valued potential that depends on the representation
theoretic data $\pi_i, i=1, \dots, 4$ as well as on the permutation $\sigma$.
We shall see a few examples below. In the case all the four external fields are
scalar, i.e.\ $\lambda_i = 0$, the potential is scalar as well and for the
trivial permutation $\sigma = \sigma_s = \textit{id}$ it reads
\begin{gather}\label{sCSHam}
V^\textit{CS}_{\hat \pi,s} (u_i)= V_{(a,b,\epsilon)}^\textit{CS}(u_i) =
V^\textit{PT}_{(a,b)}(u_1)+V^\textit{PT}_{(a,b)}(u_2)+
\frac{\epsilon(\epsilon-2)}
{8\sinh^2\frac{u_1-u_2}{2}}+\frac{\epsilon(\epsilon-2)}{8\sinh^2\frac{u_1+u_2}{2}},\\[2mm]
V^\textit{PT}_{(a,b)}(u)=\frac{(a+b)^2-\frac{1}{4}}{\sinh^2u}-\frac{ab}{\sinh^2\frac{u}{2}}\ .
\end{gather}
Here the three parameters
$a,b,\epsilon$ are given by $2a = \Delta_2-\Delta_1$, $2b = \Delta_3-\Delta_4$ and $\epsilon= d-2$.
For other elements $\sigma \in S_4$ of the permutation group, the parameters $a,b$
take different values that are given by the action of $\sigma$ on the weights. The derivation
of eq.\ \eqref{eq:LapintHam} along with the computation of the potentials for a number of
examples can be found in \cite{Schomerus:2016epl,Schomerus:2017eny,Buric:2019rms}.

\subsection{Cartan decomposition and tensor structures}

After this digression into Casimir equations and Calogero-Sutherland models
let us now come back to the study of the correlation function $G$ and in particular
the matrix $\hat\pi$ in the factorization \eqref{eq:Fpif}. In order to make further
progress we need to study the Cartan decomposition \eqref{eq:Cartan} of the elements
$g(x_i)$ in the argument of the function $F_s$ and in particular how it depends in the
insertion points $x_i$,
\begin{equation} \label{eq:CartanKPK}
g(x_i) =  e^{Ix_{21}\cdot K}e^{x_{13}\cdot P}e^{Ix_{34}\cdot K} =
k_l(x_i)\,  a(x_i)\,  k_r(x_i) \ .
\end{equation}
We will work out explicit formulas for the dependence of $k_l$ and $k_r$ on the
insertion points of the fields in $d=3,4$ dimensions below. For the moment let us
content ourselves with stating results for the factor $a(x_i)$ as well as the
dilations $d_l,d_r$ in $k_l = d_l r_l$ and $k_r = d_r r_r$. It is not too difficult
to show, see  Appendix C, that the variables $u_i$ which parametrize elements $a$
depend on $x_i$ as
\begin{eqnarray}
\label{eq:uz}
e^{u_i} & = &  1 - \frac{2}{z_i}\left(1+\sqrt{1-z_i}\right)\\[2mm]
\textit{where} \quad z_1 z_2 & = & \frac{x_{12}^2x_{34}^2}{x_{13}^2x_{24}^2} \quad , \quad
(1-z_1)(1-z_2) = \frac{x_{14}^2x_{23}^2}{x_{13}^2 x_{24}^2}\ .
\end{eqnarray}
As we suggested above, the coordinates $u_i$, are directly related to the cross ratios. In
fact, their exponentials are identical to the radial coordinates of \cite{Hogervorst:2013sma}. The relation
\eqref{eq:uz} had already been observed on a case-by-case basis in \cite{Isachenkov:2016gim,
Schomerus:2016epl,Schomerus:2017eny}. Our analysis here provides a proper derivation for all
four-point blocks, both scalar and spinning.

Similarly it is not difficult to find an explicit relation between the insertion points $x_i$
and the coordinates $\lambda_l$ and $\lambda_r$ on the dilation subgroup
\begin{equation}
d_l = d(\lambda_l) = e^{\lambda_l D} \quad , \quad d_r = d(\lambda_r) = e^{\lambda_r D}
\end{equation}
with
\begin{equation}\label{eq:lambda}
 e^{2\lambda_l} = \frac{x_{12}^2 x_{14}^2}{x_{24}^2}\frac{1}{\sqrt{(1-z_1)(1-z_2)}} \quad , \quad
 e^{2\lambda_r} = \frac{x_{14}^2}{x_{13}^2x_{34}^2}\frac{1}{\sqrt{(1-z_1)(1-z_2)}}\ .
\end{equation}
The derivation of these relatively simple formulas for $\lambda_l$ and $\lambda_r$
as well as the derivation of eq.\ \eqref{eq:uz} can be found in Appendix C. In summary we
have shown that the correlation function $G$ defined in eq.\ \eqref{eq:G} takes the form
\begin{eqnarray}
G(x_i) & = & \frac{1}{x_{12}^{2\Delta_2}x_{34}^{2\Delta_4}}
e^{2 a\lambda_l} e^{2b\lambda_r} \Theta_s(x_i) F_s \left( a(u_1,u_2)\right)\\[2mm]
& = & \label{eq:GThetaf}
\frac{1}{x_{12}^{\Delta_1+\Delta_2}x_{34}^{\Delta_3+\Delta_4}}
\left(\frac{x_{14}}{x_{24}}\right)^{2a} \left(\frac{x_{14}}{x_{13}}\right)^{2b}
[(1-z_1)(1-z_2)]^{-\frac{a}{2}-\frac{b}{2}} \, \Theta_s(x_i) f_s(u_1,u_2) \\[2mm]
& = & \label{eq:GThetapsi}
\frac{\frac12 (-1)^\frac{d-2}{2}}{x_{12}^{\Delta_1+\Delta_2}x_{34}^{\Delta_3+\Delta_4}}
\left(\frac{x_{14}}{x_{24}}\right)^{2a} \left(\frac{x_{14}}{x_{13}}\right)^{2b}
\frac{(z_1z_2)^\frac{d-1}{2}|z_1-z_2|^{-\frac{d-2}{2}} }
{[(1-z_1)(1-z_2)]^{\frac{a}{2}+\frac{b}{2}+\frac{1}{4}}} \cdot
\Theta_s(x_i) \cdot \psi_s(u_1,u_2)
\end{eqnarray}
where
\begin{equation}
\label{eq:Theta}
\Theta_s (x_i) = \hat \rho_s(x_i)\hat \pi_s(r_l(x_i),r_r(x_i)):W_P^B \mapsto W_P
\end{equation}
is a map from the space $W^B_P$ of tensor structures to the space $W_P$ of
polarisations, see also our discussion above. The rotation maps $\hat\rho_s$
and $\hat\pi_s$ were defined in eqs.\ \eqref{eq:hatrho} and \eqref{eq:hatpi}.
Recall that the index $s$ stands for $s$-channel and corresponds to the unit
element $\sigma = e \in S_4$ of the permutation group.
\smallskip

Formula \eqref{eq:GThetapsi} is the central equation of this work. It describes the
relation between four-point correlators and Calogero-Sutherland wave functions for
arbitrary fields with spin and is reminiscent of more standard decompositions of
four-point functions. Our decomposition of the function $G(x_i)$ contains three basic 
ingredients which we separated by $\cdot$ in eq.\ \eqref{eq:GThetapsi}. The first part
is a function of the insertion points which differs from the function
\begin{equation} \label{eq:DOOmega}
\Omega(x_i) = \frac{1}{x_{12}^{\Delta_1+\Delta_2}x_{34}^{\Delta_3+\Delta_4}}
\left(\frac{x_{14}}{x_{24}}\right)^{2a} \left(\frac{x_{14}}{x_{13}}\right)^{2b}
\end{equation}
in the work of Dolan and Osborn by a specific function of the cross ratios. The
second part are the tensor structures $\Theta$, i.e.\ a set of maps from the space
of tensor structures to the space of polarizations. In contrast to other approaches,
we are able to provide a universal construction through eq.\ \eqref{eq:Theta} that
allows us to construct $\Theta$ for any spin assignment through the evaluation of
certain rotation matrices. We will illustrate the usefulness of this construction
in the next sections. The final ingredient is the Calogero-Sutherland wave function
$\psi$ which takes values in the space of four-point tensor structures and replaces
the blocks $g =(g^I)$ in more conventional approaches. The functions $g^I(z,\bar z)$
satisfy Casimir equations which have been worked out in a number of examples, but depend
very much on specific choices, in particular on the basis of tensor structures. On the other
hand, with our specific choice of the other two parts, i.e.\ the function $\tilde \Omega$
and the tensor structures $\Theta$, the Casimir equations for $\psi$ are guaranteed to
take the form of a Calogero-Sutherland eigenvalue equation.
\medskip

After these comments on the formula \eqref{eq:GThetapsi} let us work out one
example right away, namely the case of scalar blocks for which the spaces $V_i \cong
\mathbb{C}$ carry a trivial representation of the rotation group and hence $\Theta_s$
is the identity map. In this case we find
\begin{eqnarray}
G(x_i) & = & \frac{1}{x_{12}^{\Delta_1+\Delta_2}x_{34}^{\Delta_3+\Delta_4}}
\left(\frac{x_{14}}{x_{24}}\right)^{2a} \left(\frac{x_{14}}{x_{13}}\right)^{2b}
[(1-z_1)(1-z_2)]^{-\frac{a}{2}-\frac{b}{2}} \, f_s(u_1,u_2)\\[2mm]
& = & \frac{1}{x_{12}^{\Delta_1+\Delta_2}x_{34}^{\Delta_3+\Delta_4}}
\left(\frac{x_{14}}{x_{24}}\right)^{2a} \left(\frac{x_{14}}{x_{13}}\right)^{2b}
\cdot \frac{(-1)^\frac{d-2}{2}}{2}
\frac{(z_1z_2)^\frac{d-1}{2}|z_1-z_2|^{-\frac{d-2}{2}} }
{[(1-z_1)(1-z_2)]^{\frac{a}{2}+\frac{b}{2}+\frac{1}{4}}}
\cdot \psi_s (u_1,u_2)\ .
\end{eqnarray}
Here we have written the result as a product of three factors which are separated by a
$\cdot$. The first factor coincides with the function \eqref{eq:DOOmega} that Dolan and
Osborn used to write the scalar four-point function $G$ as $G=\Omega g$ in terms of
a function $g = g(z,\bar z)$ of the cross ratios. The second factor denotes the gauge
transformation that was introduced in \cite{Isachenkov:2016gim} to relate blocks $g$
with Calogero-Sutherland wave functions $\psi$. The final factor $\psi$ is related to
the functions $f_s(u_1,u_2)$ in the previous line through eq.\ \eqref{eq:fomegapsi}.
Thereby we have now fully derived the relation between Calogero-Sutherland models and
scalar four-point correlators using group theory methods alone.

\section{Computation of Tensor Structures in $d=3,4$}

In the previous section we discussed how to compute the tensor structures $\Theta$
from the Cartan decomposition \eqref{eq:CartanKPK} of the group elements $g(x_i)$.
The resulting formula \eqref{eq:Theta} for $\Theta$ through representation matrices
of the elements $r_l$ and $r_r$ may still seem a bit abstract, but it is not that
difficult to work with. The precise details depend on the dimension $d$ (or rather
the value of $d\ \textit{mod}\ 8$) with dimensions $d=2n-1$ and $d=2n$ being closely
related. The first such pair $d=1$ and $d=2$ is of course trivial since all irreducible
representations of the rotation group are one-dimensional. We will therefore illustrate
our formulas with the first non-trivial pair of dimensions $d=3$ and $d=4$. In both cases,
the conformal group possesses a 4-dimensional representation which we will employ to
perform the Cartan decomposition. We begin with a few general steps that are common to
both examples before we specialize to $d=3$ dimensions in the second subsection and
$d=4$ in the third.

\subsection{Spinor representation of the conformal group}

In the two cases $d=3$ and $d=4$, the smallest non-trivial representation of the
complexified conformal algebra is 4-dimensional. Our central task is to decompose
the elements $g(x_i)$ of the conformal group as prescribed in eq.\ \eqref{eq:CartanKPK}.
We can perform this factorization in any faithful representation, including the
4-dimensional one. For the 4-dimensional conformal group, the 4-dimensional
representation is related to the realization in terms of $\textrm{SL}(2,\mathbb{H})$, see also appendix D.
The 3-dimensional conformal group $\Spin(1,4)$ is a subgroup of $\Spin(1,5)$ and hence we
can pass from the latter to the former by restricting the range of the Greek indices
$\mu,\nu=1,...,d$. These indices are raised and lowered by the flat Euclidean metric
$g_{\mu\nu} = \delta_{\mu\nu}$.

In the 4-dimensional representation, we consider a basis in which the generators of
dilations, translations and special conformal transformations read,
\begin{equation}
    D = \frac12 \begin{pmatrix}
     I  &  0\\
     0  &  -I
 \end{pmatrix},\ P_\mu = \begin{pmatrix}
     0  &  \sigma_\mu\\
     0  &  0
 \end{pmatrix},\ K_\mu = \begin{pmatrix}
     0  &  0\\
     \bar\sigma_\mu  &  0
 \end{pmatrix}. \nonumber
\end{equation}
All these matrices are written in terms of $2 \times 2$ block matrices, with $I$ denoting
the $2\times 2$ identity matrix. The matrices $\sigma_\mu$ and $\bar\sigma_\mu$ are
\begin{equation}
\sigma_\mu = (-\sigma_3,-i I,\sigma_1,-\sigma_2),\ \bar\sigma_\mu =
(-\sigma_3, i I,\sigma_1,-\sigma_2). \nonumber
\end{equation}
They are chosen such that in particular
$$ \det(a\cdot\sigma)= \det(a_\mu \sigma^\mu) = -a_\mu a^\mu = -||a||^2.$$
With these conventions, the following relations are readily seen to hold
\begin{equation}
    e^{a^\mu P_\mu} = \begin{pmatrix}
     1  &  a^\mu\sigma_\mu\\
     0  &  1
 \end{pmatrix}\quad ,\quad \  e^{a^\mu K_\mu} = \begin{pmatrix}
     1  &  0\\
     a^\mu\bar\sigma_\mu  &  1
 \end{pmatrix}. \nonumber
\end{equation}
These formulas make it easy to work out the group elements $g(x_i)$ that are defined on
the left hand side of eq.\ \eqref{eq:CartanKPK}. In the representation under consideration,
these elements become
\begin{equation}\label{specialelements}
g = e^{Ix_{21}\cdot K} e^{x_{13}\cdot P} e^{Ix_{34}\cdot K}= \begin{pmatrix}
1 + y_{13}\bar y_{34}  &  y_{13}\\
\bar y_{21} + \bar y_{34} + \bar y_{21} y_{13} \bar y_{34}  &  1 + \bar y_{21} y_{13}
\end{pmatrix},
\end{equation}
where we have introduced
\begin{equation}
\bar y_{21} = I x_{21}^\mu\bar\sigma_\mu,\ y_{13} = x_{13}^\mu\sigma_\mu,\ \bar y_{34} =
I x_{34}^\mu\bar\sigma_\mu. \nonumber
\end{equation}
This is the matrix we want to decompose into a product $k_l a k_r$. As before we split
elements $k \in K$ explicitly into dilations and rotations, i.e.\ we factorize $g$ as
\begin{equation} \label{eq:DRARD}
    g = e^{\lambda_l D} r_l a r_r e^{\lambda_r D}\ .  \nonumber
\end{equation}
In the 4-dimensional representation and with our notation using $2\times 2$ matrices, the
decomposition takes the form
\begin{equation}
    g (x_i) = \begin{pmatrix}
     e^{\frac12\lambda_l}  &  0\\
     0 & e^{-\frac12\lambda_l}
 \end{pmatrix}\, r_l(x_i)\,  \begin{pmatrix}
     C  &  S\\
     S & C
 \end{pmatrix}\, r_r(x_i)\,
      \begin{pmatrix}
     e^{\frac12\lambda_r}  &  0\\
     0 & e^{-\frac12\lambda_r}
 \end{pmatrix}, \label{eq:4dfactorization}
\end{equation}
where $\exp (\lambda_l/2)$ should be read as a $2\times 2$ matrix block, i.e.\ it is
$\exp (\lambda_l/2) {\bf 1}_2$ etc. The matrix in the middle is the 4-dimensional
representation of the factor $a(x_i)$. Its blocks are given by
\begin{equation}
    C = \begin{pmatrix}
    \cosh\frac{u_1}{2}  &  0\\
     0 & \cosh\frac{u_2}{2}
 \end{pmatrix},\ S = \begin{pmatrix}
    -\sinh\frac{u_1}{2}  &  0\\
     0 & \sinh\frac{u_2}{2}
 \end{pmatrix}. \label{eq:4da}
\end{equation}
With our choice of the dilation generator, the rotation matrices $r_l$ and $r_r$
are block diagonal, i.e.\ they read
\begin{equation}
    r_l(x_i) = \begin{pmatrix}
    L(x_i)  &  0\\
     0 & L'(x_i)
 \end{pmatrix},\ r_r(x_i) = \begin{pmatrix}
   R(x_i)  &  0\\
     0 & R'(x_i)
 \end{pmatrix}. \nonumber
\end{equation}
where $L,L',R,R'$ are $2\times 2$ rotation matrices with unit determinant. We note
that all but the middle factor $a(x)$ in the decomposition of $g$ are block diagonal.
Consequently, we can extract the coordinates $\lambda_i$ and $u_i$ from $g$ by
considering determinants of the $2\times 2$ blocks. Equating the four possible $2 \times
2$ subdeterminants in eqs.\ \eqref{specialelements} and \eqref{eq:4dfactorization}
yields the following relations
\begin{equation}
e^{4\lambda_l} = \frac{x_{12}^4 x_{13}^2 x_{14}^2}{x_{23}^2 x_{24}^2}\quad ,\quad \
e^{4\lambda_r} = \frac{x_{14}^2 x_{24}^2}{x_{13}^2 x_{23}^2 x_{34}^4}, \label{di}
\end{equation}
and
\begin{equation}
\cosh^2\frac{u_1}{2}\cosh^2\frac{u_2}{2} = \frac{x_{14}^2 x_{23}^2}
{x_{12}^2 x_{34}^2}\quad ,\quad
\sinh^2\frac{u_1}{2}\sinh^2\frac{u_2}{2} = \frac{x_{13}^2 x_{24}^2}
{x_{12}^2 x_{34}^2}, \label{torus}
\end{equation}
from which we recover the results \eqref{eq:uz} and \eqref{eq:lambda} we stated in
the previous section. Our main goal, however is to determine $r_l$ and $r_r$ now.
After moving the dilations from left hand side of eq.\ \eqref{eq:4dfactorization} to the right
we obtain
\begin{equation}
    \begin{pmatrix}
    L(x_i)CR(x_i)  &  L(x_i)SR'(x_i)\\
     L'(x_i)SR(x_i) & L'(x_i)CR'(x_i)
 \end{pmatrix} =  \begin{pmatrix}
     e^{-\frac{\lambda_l+\lambda_r}{2}}(1 + y_{13}\bar y_{34})
     &  e^{-\frac{\lambda_l-\lambda_r}{2}} y_{13}\\
     e^{\frac{\lambda_l-\lambda_r}{2}}(\bar y_{21} + \bar y_{34} + \bar y_{21} y_{13} \bar y_{34})
     &  e^{\frac{\lambda_l+\lambda_r}{2}}(1 + \bar y_{21} y_{13})
 \end{pmatrix}. \label{LRdec}
\end{equation}
The rest of our analysis is case dependent and will be carried out separately for $d=3$ and $d=4$
in the following two subsections.

\subsection{3-dimensional spinning blocks}

Calculating $\Theta_s$ is in principle straightforward, if only a bit tedious. Rotation matrices in the
3-dimensional rotation group are parametrized by three Euler angles and since the stabilizer subgroup
$B = \Spin(d-2)$ is trivial for $d=3$, we need to determine six angles in terms of the insertion points
$x_i$ by solving the equation \eqref{LRdec}. To be more precise, let us denote the generators of the
Lie algebra $\mathfrak{su}(2)$ by $\{M_{12},M_{13},M_{23}\}$. On the group we use the Euler angles
$(\phi,\theta,\psi)$ such that
\begin{equation} \label{eq:rphithetapsi}
r(\phi,\theta,\psi) = e^{-\phi M_{12}} e^{-\theta M_{23}} e^{-\psi M_{12}}.
\end{equation}
In the 4-dimensional representation the generators $M_{ij}$ are represented by
\begin{equation}\label{eq:Rmat}
 M_{23} = -\frac{i}{2} \textit{diag}(\sigma_1,-\sigma_1) \quad , \quad
 M_{13} = -\frac{i}{2} \textit{diag}(\sigma_2,\sigma_2) \quad , \quad
 M_{12} = -\frac{i}{2} \textit{diag}(\sigma_3,-\sigma_3)\ .
\end{equation}
When these matrices are inserted into eq.\ \eqref{eq:rphithetapsi}, we obtain the matrices $r_l
= r_l (\phi_l,\theta_l,\psi_l)$ and $r_r = r_r(\phi_r,\theta_r,\psi_r)$. The $2\times 2$ matrix block
$L$, for example, is given by
\begin{equation} \label{mL}
L = \begin{pmatrix}
\cos\frac{\theta_l}{2} e^{i\frac{\phi_l+\psi_l}{2}}  &  i\sin\frac{\theta_l}{2} e^{i\frac{\phi_l-\psi_l}{2}}
\\
i\sin\frac{\theta_l}{2} e^{-i\frac{\phi_l-\psi_l}{2}}  &  \cos\frac{\theta_l}{2}
e^{-i\frac{\phi_l+\psi_l}{2}}
\end{pmatrix}.
\end{equation}
and similarly for the matrix block $R$, but with angles carrying the subscript $r$ instead of $l$.
The blocks $L'$ and $R'$ are computed similarly. Note that the right hand side of eq.\ \eqref{LRdec} is written in terms of the $x_i$ already
and so are the matrix elements $C$ and $S$. Hence, eq.\ \eqref{LRdec} indeed provides us with eight
equations that allow to determine the six unknown angles in terms of $x_i$. Once these angles are
known, we can evaluate eq.\ \eqref{eq:rphithetapsi} in any representation and thereby compute $\hat
\pi_s(r_l,r_r)$ for fields of arbitrary spin.

This may sound difficult, but it can be carried out quite efficiently. We will first see how it
is done for seed blocks, i.e. when $r_l$ and $r_r$ are represented in the 2-dimensional fundamental
representation of $\mathfrak{su}(2)$. Then we employ some elementary $\textrm{SU}(2)$ representation
theory to find $\pi_s(r_l,r_r)$ for higher dimensional spin representations.

\subsubsection{Seed conformal blocks}

Let us now turn to the first example of a spinning correlator and determine its tensor
structure $\Theta_s$. The correlation function we want to look at,
\begin{equation} \label{eq:3dseed}
G^b_{\ a}(x_i) = \langle \psi^{\Delta_1}(x_1)^b \phi^{\Delta_2}(x_2) \phi^{\Delta_3}(x_2)
\psi^{\Delta_4}(x_4)_a  \rangle
\end{equation}
involves two scalar fields at $x_2$ and $x_3$ of weight $\Delta_2$ and $\Delta_3$,
respectively, along with two fermionic fields of spin $\lambda_1/2=\lambda_4/2 = 1/2$
and weight $\Delta_1$ and $\Delta_4$. In this case, the spaces $V_2$ and $V_3$ are
$1$-dimensional while $V_1$ and $V_4$ are $2$-dimensional (vectors in the spin-1/2
representation carry a Latin index $a,b...$). Hence, the space $W_P$ of
polarizations has $4$ basis elements, as does the space $W_P^B = W_P$ of tensor
structures.

For this setup we now want to compute the tensor structures $\Theta_s$. According to eq.\
\eqref{eq:Theta}, the $4\times4$ matrix $\Theta_s$ is a product of the matrix $\hat\rho_s(x_i)$
defined in eq.\ \eqref{eq:hatrho} and the matrix $\hat \pi$ which we introduced in eq.\
\eqref{eq:hatpi}. The former is very simple to calculate. Since $\pi_2$ is trivial and
$\pi_4$ is spin-1/2 representation, $\hat \rho_s$ reads
\begin{equation} \label{eq:3dhatrho}
\hat \rho_s(x_i) = \frac{1}{|x_{34}|}\left(
\begin{array}{cc}
1 & 0 \\
0 & 1 \end{array}\right)
\otimes \left(\begin{array}{cc}
x^3_{34} & -x^1_{34}-i x^2_{34} \\
x^1_{34}-i x^2_{34} & x^3_{34} \\
\end{array} \right).
\end{equation}
Calculating $\hat \pi$ is a bit more of a challenge. According to eq.\ \eqref{eq:hatpi},
$\hat \pi = \pi_{12}(r_l) \otimes \pi_{34}(r_r^{-1})$ is the tensor product of the $2\times 2$ matrices
\begin{align}
& \pi_{12}(r_l) = \begin{pmatrix}
  \cos\frac{\theta_l}{2} e^{i\frac{\phi_l+\psi_l}{2}}  &  i\sin\frac{\theta_l}{2} e^{i\frac{\phi_l-\psi_l}{2}}\\
i\sin\frac{\theta_l}{2} e^{-i\frac{\phi_l-\psi_l}{2}}  &  \cos\frac{\theta_l}{2} e^{-i\frac{\phi_l+\psi_l}{2}}
\end{pmatrix}, \label{eq:L} \\[3mm]
& \pi_{34}(r_r^{-1}) = \begin{pmatrix}
\cos\frac{\theta_r}{2} e^{i\frac{\phi_r+\psi_r}{2}}  &  -i\sin\frac{\theta_r}{2} e^{i\frac{\phi_r-\psi_r}{2}}\\
-i\sin\frac{\theta_r}{2} e^{-i\frac{\phi_r-\psi_r}{2}}  &  \cos\frac{\theta_r}{2} e^{-i\frac{\phi_r+\psi_r}{2}}
\end{pmatrix}^{-1}.\label{eq:R}
\end{align}
Through an explicit computation one may show that equation \eqref{eq:DRARD} is equivalent to
the following equation for $\hat \pi$,
\begin{equation}\label{eq:3dhatpi}
\hat\pi_s(r_l,r_r) = p(g) D_1^{-1} p(a)^{-1}.
\end{equation}
Here, $p$ is the following linear map on the space of $4\times 4$ matrices $M$,
\begin{equation}\label{map-p}
p(M)=\left(
\begin{array}{cccc}
-M_{13} & M_{12} & -M_{43} & -M_{42} \\
M_{14} & M_{11} & M_{44} & -M_{41} \\
-M_{23} & M_{22} & M_{33} & M_{32} \\
M_{24} & M_{21} & -M_{34} & M_{31} \\
\end{array}
\right).
\end{equation}
On the right hand side of eq. \eqref{eq:3dhatpi} it acts on 4-dimensional representation matrices
$g$ of elements in the conformal group as well as on the special elements $a$ whose building blocks
were stated in eq.\ \eqref{eq:4da}. The matrix $D_1$, finally, is the diagonal matrix
\begin{equation}
D_1 = \text{diag}\Big(e^{\frac{\lambda_1-\lambda_2}{2}},
e^{\frac{\lambda_1+\lambda_2}{2}},
e^{\frac{-\lambda_1-\lambda_2}{2}},
e^{\frac{-\lambda_1+\lambda_2}{2}}\Big). \nonumber
\end{equation}
To prove equation (\ref{eq:3dhatpi}), one only needs to calculate both sides. The simplicity
of the relation follows from the fact that in the 4-dimensional representation of the conformal
group, rotations are represented essentially as in the spin-1/2 representation of $\textrm{SU}
(2)$ in terms of Pauli matrices. The identity holds true for any element that admits a Cartan
decomposition. The definition of the map $p$ may come a bit ad hoc, but it has a nice group
theoretic origin that is sketched in Appendix D.

So far we thought of $\hat \pi$ as a function of six angles. But, the equation \eqref{eq:3dhatpi}
effectively represents it in terms of the insertion points $x_i$ of our four fields. To this end,
we simply insert the special elements $g$ which are constructed from the insertion points as stated
in eq.\ \eqref{specialelements}. Similarly, the matrix elements of $a$ are determined by the insertion
points through the cross ratios $u_i$, as given in eq.\ \eqref{eq:4da}. When written in terms of
the insertion points through eq.\ \eqref{di}, the diagonal matrix $D_1$ reads
\begin{equation}
D_1 = \text{diag}\Big(\sqrt{\frac{x_{12} x_{13} x_{34}}{x_{24}}},
\sqrt{\frac{x_{12} x_{14}}{x_{23} x_{34}}},\sqrt{\frac{x_{23} x_{34}}{x_{12} x_{14}}}
,\sqrt{\frac{x_{24}} {x_{12} x_{13} x_{34}}}
\Big). \nonumber
\end{equation}
This completes the construction of $\hat \pi$ and hence, along with the expression \eqref{eq:3dhatrho}
for $\hat \rho_s$, of the tensor structures $\Theta_s$ for seed blocks in $d=3$ dimensions.
\medskip

For completeness let us also briefly review the form of the corresponding Casimir equations. For seed
blocks in the Calogero-Sutherland gauge these were studied in \cite{Schomerus:2016epl}. The Hamiltonian
has block-diagonal\footnote{In the basis we use throughout this work, the first block matrix $H_1$
actually appears in the second and third row/column while $H_2$ acts on the subspace spanned by the
first and fourth basis vector} form $H=\text{diag} \{H_{1},H_{2}\}$ with the following potentials
\begin{eqnarray}\label{CasimirNewVar}
\tilde V^\textit{CS}_{1}(u_1,u_2) & =& \begin{pmatrix}
V^\textit{CS}_{(a,b,1)}(u_1,u_2) & 0 \\
0 & V^\textit{CS}_{(a,b,1)}(u_1,u_2)
\end{pmatrix}+\\[2mm]
& &\hspace*{-27mm} \begin{pmatrix}
 \frac{-1}{16}(\frac{1}{\cosh^2\frac{u_1}{2}}+\frac{1}{\cosh^2\frac{u_2}{2}}
 +\frac{2}{\cosh^2\frac{u_1-u_2}{4}}+\frac{2}{\cosh^2\frac{u_1+u_2}{4}})  &
  \frac{a+b}{4}(\frac{1}{\cosh^2\frac{u_1}{2}}-\frac{1}{\cosh^2\frac{u_2}{2}}) \\
  \frac{a+b}{4}(\frac{1}{\cosh^2\frac{u_1}{2}}-\frac{1}{\cosh^2\frac{u_2}{2}}) &
  \frac{-1}{16}(\frac{1}{\cosh^2\frac{u_1}{2}}+\frac{1}{\cosh^2\frac{u_2}{2}}
  -\frac{2}{\sinh^2\frac{u_1-u_2}{4}}-\frac{2}{\sinh^2\frac{u_1+u_2}{4}})
 \end{pmatrix}
 \notag \\[4mm]
\tilde V^\textit{CS}_{2}(u_1,u_2)  & = & \begin{pmatrix}
V^\textit{CS}_{(a,b,1)}(u_1,u_2)  & 0 \\
0 & V^\textit{CS}_{(a,b,1)}(u_1,u_2)
\end{pmatrix}+\\[2mm]
& & \hspace*{-24mm} \begin{pmatrix}
  \frac{1}{16}(\frac{1}{\sinh^2\frac{u_1}{2}}+\frac{1}{\sinh^2\frac{u_2}{2}}
  +\frac{2}{\sinh^2\frac{u_1+u_2}{4}}-\frac{2}{\cosh^2\frac{u_1-u_2}{4}}) &
  \frac{b-a}{4}(\frac{1}{\sinh^2\frac{u_1}{2}}-\frac{1}{\sinh^2\frac{u_2}{2}}) \\
  \frac{b-a}{4}(\frac{1}{\sinh^2\frac{u_1}{2}}-\frac{1}{\sinh^2\frac{u_2}{2}}) &
  \frac{1}{16}(\frac{1}{\sinh^2\frac{u_1}{2}}+\frac{1}{\sinh^2\frac{u_2}{2}}
  +\frac{2}{\sinh^2\frac{u_1-u_2}{4}}-\frac{2}{\cosh^2\frac{u_1+u_2}{4}})
\end{pmatrix}\notag
\end{eqnarray}
where \(V^{\textit{CS}}_{(a,b,1)}\) is a Calogero-Sutherland Hamiltonian, see eq.\
\eqref{sCSHam} and the $\tilde V$ differs from $V$ by a constant gauge transformation of the form
\begin{equation}  \label{eq:u}
\tilde V^\textit{CS}_i =  u V^\textit{CS}_i u^{-1} \quad \textit{with} \quad
u = \frac{1}{\sqrt{2}} \begin{pmatrix}
1 & 1\\-1 & 1 \end{pmatrix}\ .
\end{equation}
Eigenfunctions of this Hamiltonian provide us with the $s$-channel wave functions $\psi_s(u_1,u_2)$ in
eq.\ \eqref{eq:GThetapsi}. These eigenfunctions were mapped to conformal blocks in \cite{Schomerus:2016epl}.
In order to derive the transformation between blocks and Calogero-Sutherland wave functions from group
theory, we can now compare eq.\ \eqref{eq:GThetapsi} with the expansion in tensor structures that was
found in \cite{Iliesiu:2015akf}. This comparison will be carried out in the next section and we will
see that the results are in complete agreement.

\subsubsection{Tensor structures for arbitrary spins}

The tensor structures for seed blocks we constructed in the previous subsection, and in particular
the factor $\hat \pi$ given in eq.\ \eqref{eq:3dhatpi}, was so simple because $\hat\pi$ contains
the same combinations of left and right Euler angles as the matrix elements on the left hand side
of eq.\ \eqref{LRdec}. When dealing with more general spinning fields, we need to construct the
left and right Euler angles separately in terms of the insertion points. This is possible,
but the formulas are not quite as elegant as eq.\ \eqref{eq:3dhatpi}. In fact, departing from
$4\times 4$ matrix we defined in eq.\ \eqref{eq:3dhatpi} we can reconstruct the individual
factors defined in eqs.\ \eqref{eq:L} and \eqref{eq:R} as
\begin{equation} \label{eq:Lx}
 \pi_{12}(r_l) = \begin{pmatrix}
  \sqrt{\frac{\varpi_{33}}{\varpi_{11}}(\varpi_{11}\varpi_{44}-\varpi_{12}\varpi_{43})} &
  i\sqrt{\frac{\varpi_{13}}{\varpi_{31}}(1-\varpi_{11}\varpi_{44}+\varpi_{12}\varpi_{43})}\\[3mm]
  i\sqrt{\frac{\varpi_{31}}{\varpi_{13}}(1-\varpi_{11}\varpi_{44}+\varpi_{12}\varpi_{43})}
  & \sqrt{\frac{\varpi_{11}}{\varpi_{33}}(\varpi_{11}\varpi_{44}-\varpi_{12}\varpi_{43})}\\
  \end{pmatrix},
\end{equation}
and
\begin{equation} \label{eq:Rx}
  \pi_{34}(r_r^{-1}) = \begin{pmatrix}
  \sqrt{\frac{\varpi_{22}}{\varpi_{11}}(\varpi_{22}\varpi_{33}-\varpi_{13}\varpi_{42})} &
  i\sqrt{\frac{\varpi_{12}}{\varpi_{21}}(1-\varpi_{22}\varpi_{33}+\varpi_{13}\varpi_{42})}\\[3mm]
  i\sqrt{\frac{\varpi_{21}}{\varpi_{12}}(1-\varpi_{22}\varpi_{33}+\varpi_{13}\varpi_{42})}
  & \sqrt{\frac{\varpi_{11}}{\varpi_{22}}(\varpi_{22}\varpi_{33}-\varpi_{13}\varpi_{42})} \\
  \end{pmatrix}.
\end{equation}
Here $\varpi (x_i) = \hat \pi^{-1}$ denotes the inverse of the matrix $\hat \pi$ we
constructed in eq.\ \eqref{eq:3dhatpi}. Note that the matrix elements of $\varpi$ and
hence the matrix entries of the two $2\times 2$ matrices $\pi_{12}(r_l)$ and $\pi_{34}(r_r^{-1})$ are
functions of the insertion points $x_i$.

Having constructed the factors $\pi_{12}(r_l)$ and $\pi_{34}(r_r^{-1})$ for fields in the
fundamental 2-dimensional representation of $\Spin(3)$, we can now obtain these factors for all other representations with
the help of some standard group theory constructions. To fix notations let us recall that
the spin-$l$ representation of $\mathfrak{su}(2)$ is spanned by the basis vectors
$\{|l,m\rangle,m=-l,-l+1,...,l\}$. By definition, these are eigenvectors of $M_{12}$, i.e.\
they obey $M_{12}|l,m\rangle = -i m |l,m\rangle$. Matrix elements in the spin-$l$
representation are given by
\begin{equation}\label{mat-el}
t^l_{m n} (\phi,\theta,\psi) =  \langle l,m| g(\phi,\theta,\psi) | l,n\rangle =
e^{-i(m\phi+n\psi)} d^l_{m n}(\theta),
\end{equation}
where the Wigner $d$-function $d^l_{mn}$ is expressed in terms of Jacobi polynomials as
\begin{equation}\label{matrix-elements}
d^l_{m n}(\theta) = i^{m-n} \sqrt{\frac{(l+m)!(l-m)!}{(l+n)!(l-n)!}}
\Big(\sin\frac{\theta}{2}\Big)^{m-n} \Big(\cos\frac{\theta}{2}\Big)^{m+n}
P^{(m-n,m+n)}_{l-m}(\cos\theta).
\end{equation}
In particular, the spin-1/2 (fundamental) representation takes the form
\begin{equation}
\tau(\phi,\theta,\psi) = \begin{pmatrix}
\cos\frac{\theta}{2} e^{i\frac{\phi+\psi}{2}}  &  i\sin\frac{\theta}{2} e^{i\frac{\phi-\psi}{2}}\\
i\sin\frac{\theta}{2} e^{-i\frac{\phi-\psi}{2}}  &  \cos\frac{\theta}{2} e^{-i\frac{\phi+\psi}{2}}
\end{pmatrix}. \nonumber
\end{equation}
From the matrix elements of the fundamental representation, we can obtain matrix elements of any
other irreducible representation as
\begin{equation}\label{higher-reps}
t^l_{mn}(g)= (-1)^{m-n} \sqrt{\frac{(l+m)!(l-m)!}{(l+n)!(l-n)!}} \tau_{22}^{m+n} \tau_{21}^{m-n}
P^{(m-n,m+n)}_{l-m}(\tau_{11}\tau_{22}+\tau_{12}\tau_{21}).
\end{equation}
Suppose now that we want to find the tensor structures for correlators in which we insert a field
of spin $l$ at $x_1$, while keeping the field at $x_2$ to be scalar. Then the corresponding
matrix factor $\pi^l_{12}$ of $\hat\pi$ can be obtained by inserting the matrix elements
of the matrix defined in eq.\ \eqref{eq:Lx} into the previous formula. Other spin assignments
may be dealt with similarly and hence, our equations \eqref{eq:Lx}, \eqref{eq:Rx} and
\eqref{higher-reps} completely solve the problem of constructing tensor structures for all
$d=3$ dimensional spinning conformal blocks  in the Calogero-Sutherland gauge.

\subsection{4-dimensional spinning seed blocks}

Our goal in this subsection is to construct the tensor structures for the first non-trivial seed
blocks in $d=4$ dimensions. In order to do so we can follow the steps we described in the previous
subsection. In the $4$-dimensional theory, we need to determine $11$ Euler angles in total. The
rotation group $\Spin(4)$ itself is $6$-dimensional so that left and right rotations together are
parametrized by $12$ angles, but one of these angles is redundant because of the non-trivial
stabilizer subgroup $\Spin(d-2) = \Spin(2)$. To be specific, we parametrize the left and right
rotations $r_l$ and $r_r$ as
\begin{equation}
r_l = e^{\varphi^l_1 X_1} e^{\theta^l_1 Z_1} e^{\psi^l_1 X_1} e^{\varphi^l_2 X_2} e^{\theta^l_2 Z_2}
e^{\psi^l_2 X_2}\ ,\quad  r_r = e^{\varphi^r_1 X_1} e^{\theta^r_1 Z_1} e^{\psi^r_1 X_1}
e^{\varphi^r_2 X_2} e^{\theta^r_2 Z_2} e^{\psi^r_2 X_2}. \nonumber
\end{equation}
In order to reduce down to $11$ angles we impose the additional condition $\psi_2^l = -\psi_1^l$.
The symbols $X_i$ and $Z_i, i=1,2$ denote the following linear combinations of the rotation
matrices $M_{ij}$
\begin{equation}
X_1=-\frac{1}{2} (M_{12}+M_{34}),\ X_2=\frac{1}{2}(M_{12}-M_{34}),\
Z_1=-\frac{1}{2} (M_{14}+M_{23}),\ Z_2=\frac{1}{2}(M_{14}-M_{23}). \nonumber
\end{equation}
The $4 \times 4$ matrices $M_{ij}$ in turn possess a simple representation in terms of Pauli matrices.
For $M_{12}, M_{23}$ and $M_{13}$ this was spelled out already in our discussion of the 3-dimensional
theory, see eqs.\ \eqref{eq:Rmat}. The remaining matrices $M_{ij}$ are given by
\begin{equation}\label{eq:Rmat}
M_{14} = -\frac{i}{2} \textit{diag}(\sigma_1,\sigma_1) \quad , \quad
M_{24} = -\frac{i}{2} \textit{diag}(-\sigma_2,\sigma_2) \quad , \quad
M_{34} = -\frac{i}{2} \textit{diag}(\sigma_3,\sigma_3)\ .
\end{equation}
Following \cite{Schomerus:2017eny} we shall study a set of correlators that involve
the non-trivial 4-dimensional seed blocks of \cite{Echeverri:2016dun}. These turn out
to appear in the decomposition of
\begin{equation} \label{eq:4dseed}
G^b_{\ \dot a}(x_i) =
\langle \Phi_{0,0}(x_1) \Phi_{s,0}(x_2) \Phi_{0,0}(x_3) \Phi_{0,s}(x_4) \rangle\ ,
\end{equation}
where $s\in (0,1/2,1,\dots)$. Labels $(s_1,s_2)=(j_1,j_2)$ that we attached to the
operators $\Phi_{s_1,s_2}$ refer to the representation of the rotation group. We
consider the case with $s = 1/2$. The corresponding representations $V_2$ and $V_4$
are then both 2-dimensional. Their vectors are written with undotted and dotted
Latin indices, respectively. The space $W_P$ of polarizations has dimension four
and the space of $B = \Spin(2)$-invariants is 2-dimensional.
\medskip

In order to construct the map $\Theta_s$ we need to compute both $\hat \rho_s$ and $\hat \pi$.
The former is given by
\begin{equation}\label{eq:4dhatrho}
\hat \rho_s(x_i) = \frac{1}{|x_{12}||x_{34}|}\left(
\begin{array}{cc}
x_{21}^4-i x_{21}^3 & x_{21}^2-i x_{21}^1 \\
-x_{21}^2-i x_{21}^1 & x_{21}^4+i x_{21}^3 \end{array}\right)
\otimes \left(
\begin{array}{cc}
x_{34}^4-i x_{34}^3 & x_{34}^2+i x_{34}^1 \\
-x_{34}^2+i x_{34}^1 & x_{34}^4+i x_{34}^3 \end{array}\right)\ .
\end{equation}
This is obtained by evaluating the rotation matrix $s_{x} s_{e_d}$ in the 2-dimensional
representations $(1/2,0)$ and $(0,1/2)$ of the rotation group, see eq.\ \eqref{eq:hatrho}.
Calculating $\hat \pi$ is a bit more involved. With Euler angles introduced as described
above, the left and right representations of the rotation generators are given by
\begin{align*}
& \pi_{12}(r_l) = \begin{pmatrix}
\cos\frac{\theta_2^l}{2} e^{-i\frac{\phi_2^l+\psi_2^l}{2}}  &
i\sin\frac{\theta_2^l}{2} e^{-i\frac{\phi_2^l-\psi_2^l}{2}}\\
i\sin\frac{\theta_2^l}{2} e^{i\frac{\phi_2^l-\psi_2^l}{2}}  &
\cos\frac{\theta_2^l}{2} e^{i\frac{\phi_2^l+\psi_2^l}{2}}
\end{pmatrix},\\[2mm]
& \pi_{34}(r_r) = \begin{pmatrix}
\cos\frac{\theta_1^r}{2} e^{-i\frac{\phi_1^r+\psi_1^r}{2}}  &
i\sin\frac{\theta_1^r}{2} e^{-i\frac{\phi_1^r-\psi_1^r}{2}}\\
i\sin\frac{\theta_1^r}{2} e^{i\frac{\phi_1^r-\psi_1^r}{2}}  &
\cos\frac{\theta_1^r}{2} e^{i\frac{\phi_1^r+\psi_1^r}{2}}
\end{pmatrix},
\end{align*}
where $\psi_2^l = -\psi_1^l$. We need to calculate the tensor product $\hat \pi$  of
these two matrices as a function of the insertion points $x_i$ to obtain the main
building block for the desired tensor structure. As in the previous section the
resulting expression is surprisingly simple. It requires to introduce the following
linear map $q:M_{2\times2}\times M_{2\times 2} \xrightarrow{}M_{2\times 4}$ that
sends a pair of $2\times 2$ matrices $M,N$ to a rectangular $2\times 4$ matrix
of the form
\begin{equation}
q(M,N)= \begin{pmatrix} r(M)\\
r(N)\\
\end{pmatrix},\ r(M)=\begin{pmatrix}
M_{12} & M_{11} & M_{22} & M_{21} \\
\end{pmatrix}. \nonumber
\end{equation}
From the four $2\times 2$ matrix blocks of the group element $g$ in the
$4$-dimensional representation we can construct a pair of $2\times 2$ matrices
$M = M_2$ and $N = M_1$ as
\begin{equation}\label{Mis}
M_i = \sinh ^2\frac{u_i}{2}D B^{-1} A - \cosh ^2\frac{u_i}{2}C
= \sinh^2\frac{u_i}{2}y_{13}^{-1}-\bar y_{21}-\bar y_{34} - \bar y_{21} y_{13}\bar y_{34}.
\end{equation}
The second equality applies to special elements $g(x_i)$ that we assigned to a set
of insertion points in eq.\ \eqref{specialelements}. In terms of these two matrices
one can now compute $\hat \pi$ as
\begin{equation}
\hat \pi(r_l,r_r)^{-1} = \mathcal{P}(\pi_{12}(r_l)\otimes\pi_{34}(r_r^{-1}))^{-1} =
\frac{2 e^{\frac{\lambda_1-\lambda_2}{2}}} {\cosh u_1-\cosh u_2}\begin{pmatrix}
\sinh\frac{u_1}{2} & 0 \\
0 & \sinh\frac{u_2}{2}\\
\end{pmatrix} q(M_2,M_1).
\label{main4}
\end{equation}
Here, $\mathcal{P}$ is the projector to the space of $B$-invariants
\begin{equation}
\mathcal{P} = \begin{pmatrix}
0 & 1 & 0 & 0\\
0 & 0 & 1 & 0\\
\end{pmatrix}, \nonumber
\end{equation}
see \cite{Schomerus:2017eny} for its derivation. Note that the matrices $\hat\pi$ and $\hat\pi^{-1}$
are not square. When we invert the $2\times 4$ matrix $\hat \pi^{-1}$ defined in eq.\ \eqref{main4}
we are instructed to build a $4\times 2$ matrix $\hat \pi$ such that
\begin{equation}
\hat\pi(r_l,r_r)^{-1}\hat\pi(r_l,r_r)=I_2. \nonumber
\end{equation}
Again, the identity \eqref{main4} holds for any element of the conformal group that has a Cartan decomposition,
as can be checked by calculating both sides. The application to $g(x_i)$ is obtained by substituting
eq.\ (\ref{Mis}) into eq.\ (\ref{main4}).
\medskip

The Calogero-Sutherland potential for the case under consideration was constructed in \cite{Schomerus:2017eny},
and reads
\begin{eqnarray}\label{Spin1halfCasimir}
& & \hspace*{-5mm}\tilde V^\textit{CS}_{\frac{1}{2}}(u_1,u_2)
=\begin{pmatrix}
V^\textit{CS}_{(a,b,2)}(u_1,u_2) -\frac{1}{8} & 0 \\
0 & V^\textit{CS}_{(a,b,2)}(u_1,u_2)-\frac{1}{8}
\end{pmatrix}+\\[2mm]
& &  \frac{1}{16}\begin{pmatrix}
\frac{1}{\sinh^2\frac{u_1}{2}}+\frac{1}{\sinh^2\frac{u_2}{2}}+\frac{4}{\sinh^2\frac{u_1-u_2}{4}}-
\frac{4}{\cosh^2\frac{u_1+u_2}{4}} & 4(b-a)\left(\frac{1}{\sinh^2\frac{u_1}{2}}-\frac{1}{\sinh^2\frac{u_2}{2}}\right) \\
4(b-a) \left( \frac{1}{\sinh^2\frac{u_1}{2}}-\frac{1}{\sinh^2\frac{u_2}{2}}\right) & \frac{1}{\sinh^2\frac{u_1}{2}}+
\frac{1}{\sinh^2\frac{u_2}{2}}+\frac{4}{\sinh^2\frac{u_1+u_2}{4}}-\frac{4}{\cosh^2\frac{u_1-u_2}{4}}
\end{pmatrix}\ . \notag
\end{eqnarray}
where once again $\tilde V$ is related to the potential $V$ by the same constant gauge transformation
$u$ we introduced in eq.\ \eqref{eq:u}.
\smallskip

The tensor structures $\Theta_s \sim \hat\rho_s \hat\pi$ we computed above resembles the tensor structures
constructed in \cite{Echeverri:2016dun}. In \cite{Schomerus:2017eny} a map $\chi$ was found that
sends the conformal blocks of \cite{Echeverri:2016dun} to the Calogero-Sutherland wave functions $\psi_s$.
One may show that the composition of this map with the tensor structures of \cite{Echeverri:2016dun}
reproduces the tensor structures $\Theta_s$ we have constructed in this subsection, see next section.

Recall that in the 3-dimensional theory we discussed how to obtain tensor structures for any spin assignments
in the correlator. The construction proceeded in two steps. First we factorized the construction of
$\hat \pi$ in eq.\ \eqref{eq:3dhatpi} for the seed block into explicit expressions for the
fundamental left and right matrices in terms of insertion points, see eqs.\ \eqref{eq:Lx} and
\eqref{eq:Rx}. Then, in a second step, we used standard $\textrm{SU}(2)$ group theory to build
left and right representation matrices in any other representation. Here we could do something
very similar except that the information encoded in eq.\ \eqref{main4} is not sufficient to
reconstruct all $11$ Euler angles and hence to reconstruct the left and right rotation matrices.
In fact, the linear map $\hat \pi$ given in eq.\ \eqref{main4} only contains $2\times 4 = 8$
matrix elements and hence we do not have enough equations to solve for the $11$ angles. In
order to obtain the missing equations one needs to compute $\hat \pi$ for one other spin
assignment. Once the left and right rotations are fixed in terms of the insertion points of
the four fields, one can proceed as in the 3-dimensional case since the 4-dimensional rotation
group $\SO(4)$ is a double cover of $\SO(3) \times \SO(3)$. Hence representation matrices in
higher spin representations can be computed from those in the two fundamental representations
in the same way as we explained at the end of the previous subsection. So, in principle it is
not much more complicated to obtain tensor structures for arbitrary spin assignments in $d=4$
dimensions.

\section{Comparison with other Approaches}

In this section, we want to compare our tensor structures in Calogero-Sutherland gauge with the ones that were
constructed in the literature previously, most notably in \cite{Iliesiu:2015akf} on 3-dimensional seed blocks
and in \cite{Echeverri:2016dun} for the 4-dimensional case. Conformal blocks in these two different gauges had
previously been related in \cite{Schomerus:2016epl} and \cite{Schomerus:2017eny}, respectively, through a
comparison of the relevant Casimir equations. Now that we have a systematic theory of tensor structures for
the Calogero-Sutherland gauge, it is possible to also compare the tensor structures with the ones in the
literature, at least for the cases that have been worked out. It is a nice consistency check to see that
the results match.

In conventional approaches to spinning conformal blocks, one chooses some more or less natural or simple
set of tensor structures for each case at hand, i.e.\ a given assignment of spins, and then expands the
correlator in these tensor structures,
\begin{equation} \label{Gdecstandart}
G^a(\pi_i;x_i) = \Omega(\Delta_i;x_i) \sum_I t^a_I(x_i)  g^I(\pi_i;z,\bar z) \ .
\end{equation}
Here, $\Omega$ is the usual scalar function \eqref{eq:DOOmega} that was introduced by Dolan and Osborn,
the superscript $a$ runs through a basis of polarizations and $I$ labels some basis of tensor structures.
We shall think of $T = (t^a_I)$ as a matrix of tensor structures. As in the text above, it maps the space
of tensor structures to the space of polarizations. The tensor structures $T$ are chosen such that
the functions $g^I(z,\bar z)$ depend on cross ratios only. Once the tensor structures are fixed, one
can derive Casimir equations for the associated conformal blocks $g^I$. Of course the form of the
Casimir equations depends on the choice of the tensor structures.

In contrast to this standard theory, our construction of tensor structures $\Theta_s$ is systematic
and canonical. This implies that the Casimir equations are canonical as well, i.e.\ they always come
out to be of Calogero-Sutherland form. The decomposition of spinning correlators in terms of
Calogero-Sutherland wave functions was given in eq.\ \eqref{eq:result}. This should be compared
with the decomposition \eqref{Gdecstandart}, i.e.\ symbolically
\begin{equation}
\Omega \, T \, \chi = \tilde \Omega \, \Theta\,
\end{equation}
where $\chi= (\chi^{I}_{J})$ is a transformation in the space of tensor structures that relates the
blocks $g^I$ to the Calogero Sutherland wave functions $\psi^J$. We will now compute this map $\chi$
for seed blocks in $d=3$ and $d=4$ dimensions and compare with the results in \cite{Schomerus:2016epl}
and \cite{Schomerus:2017eny}. Throughout this section we shall omit the subscript $s$ since all
quantities are in the $s$-channel.

\subsection{3-dimensional spinning blocks}

Let us begin the discussion of the 3-dimensional case by reviewing the construction of tensor
structures for the only non-trivial seed blocks in \cite{Iliesiu:2015akf}. The constructions in
that work are performed in the Minkowski space with the metric
\begin{equation}
g_{\mu\nu} = \text{diag}(-1,1,1). \nonumber
\end{equation}
Greek indices $\mu,\nu...$ are raised and lowered with this metric. Greek indices $\alpha,\beta...$
from the beginning of the alphabet are raised and lowered with the Sp$(2)$ symplectic form according to
\begin{equation}
\Omega_{\alpha\beta} = \begin{pmatrix}
0 & 1\\
-1 & 0\\
\end{pmatrix},\ \psi_{\alpha} = \Omega_{\alpha\beta}\psi^\beta. \nonumber
\end{equation}
A vector $x^\mu$ is made into a $2\times2$ matrix with the help of 3-dimensional gamma matrices
\begin{equation}
(\gamma^\mu)^\alpha_{\ \beta} = (i\sigma_2,\sigma_1,\sigma_3), \nonumber
\end{equation}
where $\sigma_i$ are the Pauli matrices. We write
\begin{equation}
x^\alpha_{\ \beta} = x_\mu (\gamma^\mu)^\alpha_{\ \beta}\ .
\end{equation}
The correlation function \eqref{eq:3dseed} of two scalar and two spinor fields then assumes
the form
\begin{equation}
    \langle \psi^\alpha_1(x_1) \varphi_2(x_2) \varphi_3(x_3) \psi_4^\beta(x_4) \rangle = \Omega(x_i)
    \sum_{I=1}^4 t^{\alpha\beta}_I g^I(z,\bar z), \nonumber
\end{equation}
where the index $I=1,\dots,4$ runs over the following four four-point tensor structures
\begin{align*}
   & t_1^{\alpha\beta} = i\frac{(x_{14}i\sigma_2)^{\alpha\beta}}{|x_{14}|}, &&
   t_2^{\alpha\beta} = -i \frac{(x_{12}x_{23}x_{34}i\sigma_2)^{\alpha\beta}}{|x_{12}||x_{23}||x_{34}|},\\[2mm]
   & t_3^{\alpha\beta} = i\frac{(x_{12}x_{24}i\sigma_2)^{\alpha\beta}}{|x_{12}| |x_{24}|},
   && t_4^{\alpha\beta} = i\frac{(x_{13}x_{34}i\sigma_2)^{\alpha\beta}}{|x_{13}||x_{34}|}.
\end{align*}
and $\Omega(x_i)$ is the function \eqref{eq:DOOmega} introduced by Dolan and Osborn. The matrix $T = (t^{\alpha\beta}_I)$
is the most nontrivial factor that relates the correlation function $G$ with the conformal blocks $g^I$.
It bears some resemblance with the tensor structures $\Theta = \Theta_s$ we computed in eq.\ Section 4.2, and most
notably with the matrix $p(g(x_i))$ in eq.\ \eqref{eq:3dhatpi}. More precisely one can see that
\begin{equation} \label{eq:3dChi}
c\,  T\,  \chi =  \hat \rho\,  p(g)\,  D_1^{-1} \, p(a)^{-1}
\end{equation}
where $c = \Omega/\tilde \Omega$
and the matrix $\chi$ takes the form
\begin{equation}
\chi =\left(
\begin{array}{cccc}
0 & A(\chi_1-\chi_2) & A(-\chi_1-\chi_2) & 0 \\
0 & A(-\chi_1-\chi_2) & A(\chi_1-\chi_2) & 0 \\
B(\chi_3-\chi_4) & 0 & 0 & B(\chi_3+\chi_4) \\
B(-\chi_3-\chi_4) & 0 & 0 & B(-\chi_3+\chi_4) \\
\end{array}
\right). \nonumber
\end{equation}
with $A=-i/2$ and $B = -1/2$.\footnote{The comparison of Casimir equations that was used in \cite{Schomerus:2016epl}
to determine $\chi$ does not determine the numerical factors $A$ and $B$. The map that was spelled out in that work
corresponds to $A=1/\sqrt{2}=B$.} The functions $\chi_i(u_1,u_2)$ are defined in (A.10) of \cite{Schomerus:2016epl}, with $x=u_1,\ y=u_2$. Let us note that in checking eq.\ \eqref{eq:3dChi} it is advantageous to bring the factor
$p(a)$ to the left hand side and use
\begin{equation}
c\,  \chi\,  p(a(x_i)) = \left(
\begin{array}{cccc}
0 & \sqrt{\frac{x_{14}x_{23}}{x_{12}x_{34}}} & 0 & 0 \\
0 & 0 & \sqrt{\frac{x_{14}x_{23}}{x_{12}x_{34}}} & 0 \\
0 & 0 & 0 &   \sqrt{\frac{x_{13}x_{24}}{x_{12}x_{34}}} \\
 \sqrt{\frac{x_{13}x_{24}}{x_{12}x_{34}}} & 0 & 0 & 0 \\
\end{array}
\right). \nonumber
\end{equation}
This concludes our discussion of tensor structures for 3-dimensional seed blocks. The tensor structures for
other blocks have not been stated in the literature.

\subsection{4-dimensional spinning blocks}

Let us begin the discussion of the 4-dimensional case by reviewing the construction of tensor
structures for seed blocks in \cite{Echeverri:2016dun,Cuomo:2017wme}. It is set in Minkowski
signature with the metric
\begin{equation}
g_{\mu\nu} = \text{diag}(-1,1,1,1). \nonumber
\end{equation}
Greek indices from the second half of the alphabet are raised and lowered with this metric $g$.
Indices $\alpha,\beta...$ from the beginning of the Greek alphabet label the basis of the
$2$-dimensional representation $(1/2,0)$ of $\SO(1,3)$. Similarly, the dotted indices $\dot\alpha,
\dot\beta...$ enumerate a basis in the representation $(0,1/2)$. These are raised and lowered
with the Levi-Civita symbol according to
\begin{equation}
\psi_\alpha = \varepsilon_{\alpha\beta}\psi^\beta,\  \varepsilon_{\alpha\beta} =
\begin{pmatrix}
0 & -1\\
1 & 0\\
\end{pmatrix}. \nonumber
\end{equation}
The same formulas hold for the dotted indices. The vector representation is equivalent to the
tensor product $(1/2,0)\otimes(0,1/2)$. This equivalence can be realised explicitly with the
help of $\sigma$-matrices
\begin{equation}
x_{\alpha\dot\beta} = x_\mu\sigma^\mu_{\alpha\dot\beta},\ \sigma^\mu_{\alpha\dot\beta} =
(-I,\sigma^i). \nonumber
\end{equation}
Further we write $(\bar\sigma^\mu)^{\dot\alpha\beta} = (-I,-\sigma^i)$ and define the corresponding
$\bar x^{\dot\alpha\beta}$ in the obvious way.

With this preparation let us now look at the correlation function \eqref{eq:4dseed}.
It decomposes over the tensor structures introduced above as\footnote{Following the
  conventions in \cite{Echeverri:2016dun} we label seed blocks
by an integer $p = 2s$ rather than the spin $s$ itself.}
\begin{equation}
G^{\dot\alpha\beta} (x_i) = \mathcal{K}_4(x_i)\sum_I T^{\dot\alpha\beta}_I(x_i)\,  g^I(U,V) =
\mathcal{K}_4(x_i) \sum_{e=0}^{2s} g_e^{(2s)}(U,V) I^{e}_{42} J^{p-e}_{42,31} , \nonumber
\end{equation}
where the scalar prefactor $\mathcal{K}_4$ is defined by
\begin{equation}
\mathcal{K}_4 = \Omega \sqrt{\frac{x_{13}}{x_{12} x_{24} x_{34}}} \nonumber
\end{equation}
and the tensors $I, J$ on the right hand side take the form
\begin{align*}
& I_{ij}^{\dot\alpha\beta} = x^\mu_{ij} \bar\sigma_\mu^{\dot\alpha\beta},\\[2mm]
& J_{ij,kl}^{\dot\alpha\beta} = \frac{-1}{x^2_{kl}}\Big( [x_{ik},x_{jl}]^\mu +
[x_{jk},x_{il}]^\mu +[x_{ij},x_{lk}]^\mu - 2i \varepsilon^{\mu\nu\rho\sigma}
x_{ik\nu} x_{lj\rho} x_{lk\sigma}\Big) \bar\sigma_\mu^{\dot\alpha\beta},
\end{align*}
with $[x,y]^\mu = x^2 y^\mu - y^2 x^\mu$. The second tensor can be written more
compactly as
\begin{equation}
J_{ij,kl}^{\dot\alpha\beta} =  \frac{2}{x_{kl}^2} \bar x_{ik}^{\dot\alpha\gamma}
(x_{kl})_{\gamma\dot\delta} \bar x_{lj}^{\dot\delta\beta}. \nonumber
\end{equation}
The case we analysed in Section 4.3 corresponds to $2s = p = 1$ and hence the
correlation function has the form
\begin{equation}
G^{\dot\alpha\beta}(x_i) = \mathcal{K}_4(x_i) \Big( g_0^{(1)}(U,V) J^{\dot\alpha\beta}_{42,31}
+ g_1^{(1)}(U,V) I^{\dot\alpha\beta}_{42} \Big) . \nonumber
\end{equation}
involving two tensor structures, in agreement with the dimension of the space
$W_P^B$ of tensor structures in our analysis.
\smallskip

In order to compare the tensor structures from \cite{Echeverri:2016dun} with our tensor
structures given in eqs.\ \eqref{eq:4dhatrho} and \eqref{main4} we note that the two
discussions of the seed correlators use a different basis in the space of polarizations.
Comparing the conventions in \cite{Cuomo:2017wme} and \cite{Schomerus:2017eny} one can
see that the basis transformation is mediated by the following matrix
\begin{equation}
M^{b\ \beta}_{\ \dot a\ \dot \alpha}=\frac{i}{2}\begin{pmatrix}
1 & 1 & -1 & -1\\
-1 & -1 & -1 & -1\\
1 & -1 & -1 & 1\\
-1 & 1 & -1 & 1\\
\end{pmatrix}. \nonumber
\end{equation}
Taking this necessary change of basis into account we have to prove that
\begin{equation}\label{eq:MOTcOT}
M \Omega \, T \, \chi = \tilde \Omega \, \Theta\, .
\end{equation}
It is not difficult to check that this equation is indeed satisfied with $\chi$
given by
\begin{equation}
\chi = \frac{(\coth{\frac{u_1}{2}}\coth{\frac{u_2}{2}})^{-a-b-\frac12}}
{(\cosh{u_1}-\cosh{u_2})^2} \begin{pmatrix}
\sinh{\frac{u_1}{2}} & \sinh{\frac{u_2}{2}}\\
2\sinh^{-1}\frac{u_1}{2} & 2\sinh^{-1}\frac{u_2}{2} \\
\end{pmatrix}. \nonumber
\end{equation}
This coincides with the map between Calogero-Sutherland eigenfunctions and conformal blocks that was found
in \cite{Schomerus:2017eny} based on the comparison of the Casimir differential equations. The easiest way
to verify eq.\ \eqref{eq:MOTcOT} passes through the following formula
\begin{equation}
c\chi\hat\pi^{-1} = \frac{e^{\frac{\lambda_1-\lambda_2}{2}}
\sinh{\frac{u_1}{2}}\sinh{\frac{u_2}{2}}}{(\cosh u_1 - \cosh u_2)^2}
\begin{pmatrix}
2\sinh^2\frac{u_1}{2}\sinh^2\frac{u_2}{2} r_1 - (\sinh^2\frac{u_1}{2}+\sinh^2\frac{u_2}{2}) r_2\\
2(\sinh^2\frac{u_1}{2}+\sinh^2\frac{u_2}{2}) r_1-4r_2\\
\end{pmatrix},\nonumber
\end{equation}
where $c$ is the ratio $c = \Omega/\tilde \Omega$ we defined in the previous subsection and
we introduced
\begin{equation}
r_1 = r(y_{13}^{-1}),\ r_2 = r(\bar y_{21}+\bar y_{34}+\bar y_{21} y_{13}\bar y_{34}).\nonumber
\end{equation}
With a bit of rewriting it follows that
\begin{equation}
c\frac{\mathcal{K}_4}{\Omega}\chi \hat\pi^{-1} = \frac{x_{12}x_{13}}{x_{24}^2}
\frac{1}{4((1+U-V)^2-4U)}\begin{pmatrix}
2r_1 + (1+U-V) r_2\\
-2(1+U-V)r_1-4Ur_2\\
\end{pmatrix}. \nonumber
\end{equation}
Thus, we need to show
\begin{equation}
\begin{pmatrix}
r_1\hat\rho^{-1}\\
r_2\hat\rho^{-1}\\
\end{pmatrix}M\begin{pmatrix}
T_1 & T_2\\
\end{pmatrix}=\frac{x_{24}^2}{x_{12}x_{34}}\begin{pmatrix}
4U & 1+U-V\\
-2(1+U-V) & -2\\
\end{pmatrix}, \nonumber
\end{equation}
and this is easily confirmed, supporting our results for tensor structures of
4-dimensional seed blocks. This concludes our comparison of tensor structures for
3- and 4-dimensional seed blocks. Even though the steps were a bit technical, we
hope that our discussion helps to better appreciate the similarities of our group
theoretic approach with more standard constructions of tensor structures. While
the resulting matrix factors can be related by some simple transformations $\chi$,
our approach does provide an algorithm for arbitrary spin assignments and it
ensures that the associated Casimir equations take the form of eigenvalue
equations for some spinning Calogero-Sutherland Hamiltonian.

\section{Conclusions and Outlook}

In this work we employed conformal group theory to explicitly embed the theory of Calogero-Sutherland
models and their wave functions into conformal field theory. In the case relevant for four-point
functions of spinning fields, Calogero-Sutherland wave functions depend on two variables $u_i, i =
1, 2$, which can be considered as coordinates on a particular 2-dimensional abelian subgroup of
the conformal group, as described in eq.\ \eqref{adef}. The precise relation to the usual cross
ratios $z$ and $\bar z$ had been inferred from the comparison between Casimir equations and
eigenvalue equations for Calogero-Sutherland models in \cite{Isachenkov:2016gim}. Here we gave
a first principle derivation. Let us note that the (exponentials of the) coordinates $u_i$ had appeared
as {\em radial coordinates} in the work of Hogervorst and Rychkov \cite{Hogervorst:2013sma} before their
relevance for the relation with Calogero-Sutherland models had been appreciated. According to
the general discussion in Section 2, any function of the two variables $u_i$ can be  extended
to a function of $4d$ variables $x_i$ that satisfies all conformal Ward identities. This
extension only depends on the choice of a channel, i.e.\ an element $\sigma$ of the permutation
group $S_4$. The map that provides the extension from the cross ratios to the set of insertion
points naturally splits into two factors, one that is associated with the subgroup $\SO(1,1)$
of dilations and another associated with the rotation group. The former is a scalar function
$\tilde \Omega$ that had also been inferred before from the comparison of Casimir and
Calogero-Sutherland equations \cite{Isachenkov:2016gim} and was now computed independently. The
latter function $\Theta$ is matrix valued. More specifically it maps the space of four-point tensor
structures to the space of polarizations of the four fields. In analogy to the standard
treatments of spinning correlation functions we referred to $\Theta$ as {\em tensor structures}.
The main new result of this work was to compute $\Theta$ by means of the Cartan decomposition
of the conformal group. This was first described in general in Section 3 and then carried out
explicitly for a few cases in Section 4. While our central formula \eqref{eq:result2} resembles
similar decompositions in standard treatments of spinning correlators, it is entirely group
theoretical in nature and very universal since all Casimir equations in conformal field theory
possess a Calogero-Sutherland formulation. By now the latter is known for quite a few cases
in which Casimir equations have never been worked out, see also below.
\medskip

As we stressed in the previous paragraph, the extension of a function $\psi$ of the two
variables $u_i$ to a function of $4d$ variables $x_i^\mu$ depends on the choice of a channel,
i.e. a permutation $\sigma \in S_4$ of four points in the $d$-dimensional space. Passing
(crossing) from one channel $\sigma$ to another $\sigma'$ involves the following crossing
matrices
\begin{equation} \label{XiinvXi}
\Xi^{\hat \pi,+}_{\sigma'}(x_i)\  \Xi^{\hat \pi}_\sigma(x_i):
 W^B_P \mapsto W^B_P
\end{equation}
where $\Xi^+:W_P \mapsto W_P^B$ denotes a left-inverse of $\Xi$. One may think of the
product \eqref{XiinvXi} as a square matrix of cross ratios. For scalar fields of equal
weight $\Delta_i = \Delta$ the transition element from the $s$- to the $t$-channel, i.e.
with $\sigma = \sigma_s = \textit{id}$ and $\sigma' = \sigma_t = (24)$, reads
\begin{equation}\label{XiinvXiscalar}
\Xi^{\Delta,+}_t(x_i)\  \Xi^{\Delta}_s(x_i) =
\left[ \frac{z_1z_2}{(1-z_1)(1-z_2)}\right]^{\frac{d}{2}- \frac{1}{4} - \Delta}  =
\left[ \frac{U}{V}\right]^{\frac{d}{2}- \frac{1}{4} - \Delta} \ .
\end{equation}
Of course for scalar fields this result is well known, except that we display it here
in Calogero-Sutherland gauge rather than the usual conformal one. Since the transition
matrices \eqref{XiinvXi} are required to spell out the crossing symmetry constraints for
functions of cross ratios it would be interesting to compute these products for spinning
blocks. We have not done so in general, but can quote at least one result that applies to
the $s=1/2$ seed blocks in $d=4$ dimensions. In this case, the crossing matrix \eqref{XiinvXi}
is easy to compute from the explicit expressions \eqref{main4} for $\hat \pi$ and
eq.\ \eqref{eq:4dhatrho} for $\hat \rho$. The result is
\begin{equation}
(\hat\pi_t)^{-1}\hat\rho_t^{-1}\hat\rho_s\hat\pi_s =
-i\Big(\cosh\frac{u_1}{2}\cosh\frac{u_2}{2}\Big)^{-1/2}\begin{pmatrix}
\cosh\frac{u_1}{2} & 0\\
0 & \cosh\frac{u_2}{2}\\
\end{pmatrix}.
\end{equation}
Here the maps $\hat\rho_t$ and $\hat \pi_t$ have been obtained from our expressions for
$\hat \rho_s$ and $\hat \pi_s$ by exchanging $x_1$ and $x_3$. The result for the transition
matrix is surprisingly simple in this case. It would clearly be of interest to compute such
crossing matrices for more general spinning blocks.
\medskip

One of the main motivations for this work was to treat setups in which the tensor
structures have not been computed with any other techniques. This applies in particular
to the case of superblocks for external operators in long, non-BPS multiplets. For a large
class of superconformal symmetries, the Calogero-Sutherland type Casimir equations were
recently obtained in \cite{Buric:2019rms}. Explicit computations of the corresponding wave functions
$\psi$ for $\mathcal{N}=1$ superconformal field theories in $d=4$ dimensions are being worked out
in \cite{Buric2}. In order to relate these wave functions to the correlators $G$ it remains to
determine the supersymmetric analogues of $\tilde \Omega$ and $\Theta$. We will address this
problem in future work.
\smallskip

Another important context in which tensor structures are poorly studied concerns  defect
correlators, i.e. correlation functions that involve one or more non-local operators like
line or surface defects etc. Once again, the Calogero-Sutherland models for the relevant
wave functions $\psi$  are known from \cite{Liendo:2018ukf}, at least in the scalar case.
From this work on may also infer the analogue of the factor $\tilde \Omega$. Extensions 
to spinning defect correlators and their blocks have been initiated in \cite{Lauria:2018klo,
Kobayashi:2018okw} and it would be interesting to extend the Calogero-Sutherland approach 
including our group theoretic construction of the relevant tensor structures $\Theta$ to 
spinning defect correlators. 
\bigskip \bigskip

\noindent
{\bf Acknowledgement:} We are indebted to Valya Petkova for an interesting conversation that
triggered this work. We also thank Giovanni Felder, Madalena Lemos, Pedro Liendo, Slava Rychkov,
Aleix Gimenez-Grau and in particular Zhenya Sobko for discussions and comments. This work was
supported in parts by the Deutsche Forschungsgemeinschaft (DFG, German Research Foundation)
under Germany‘s Excellence Strategy – EXC 2121 \textit{Quantum Universe} – 390833306.  This
research has also received funding from the European Research Council (ERC) under the European
Union’s Horizon 2020 research and innovation program (QUASIFT grant agreement 677368).

\appendix

\section{Verification of Ward identities}

The aim of this appendix is to show that the function constructed on the left hand side eq.\ \eqref{maine} indeed
satisfies all the Ward identities a four-point correlation function of spinning fields has to satisfy. In Section 2
we discussed the Ward identities for infinitesimal conformal transformations. Of course, these all integrate to
global conformal transformations. Here we shall establish these global Ward identities. Most of our notations were
introduced in Section 2 already, but we need one more ingredient now concerning the action of global conformal
transformations. An element $g \in G$ sends a point $x$ in Euclidean space to $gx$. The differential of $g$ at
$x$ will be denoted by $dg_x$ - it sends tangent vectors at $x$ to tangent vectors at $gx$. Since the spacetime
is just the flat space, we can identify the tangent space at any point with the spacetime itself. After this is
done, the statement that $g$ is a conformal transformation means precisely that, at any point $x$, $dg_x$ belongs
to the subgroup $K = \SO(1,1) \times \Spin(d)$ that is generated by dilations and rotations of the tangent space. In
case $g$ is a translation, $dg_x$ is trivial. Since dilations and rotations are linear maps, their differentials
are simply the maps themselves (in the obvious sense). For the proof that follows, we will need the differential
of the Weyl inversion, which was derived in Section 2.1: $dw_x = x^{-2}s_{e_d} s_x $. The global version of the
result stated in Section 2 reads
\medskip

\noindent
{\bf Theorem:} \textit{Let $F:G\xrightarrow{}W_P$ be a $K \times K$ covariant function satisfying the following
covariance law
\begin{equation} \label{eq:covariance}
F(k_l g k_r) = \pi_{12}(k_l) \otimes \pi_{34}(k_r^{-1}) F(g), \nonumber
\end{equation}
and define a new function by the right hand side of eq.\ \eqref{maine}, i.e. through the expression
\begin{equation}
G(x_i) = \frac{1}{x_{12}^{2\Delta_2} x_{34}^{2\Delta_4}} \Big(1\otimes\pi_2(s_{x_{12}}s_{e_d})\otimes1
\otimes\pi_4(s_{x_{34}}s_{e_d})\Big) F(e^{Ix_{21}\cdot K}e^{x_{13}\cdot P}e^{I x_{34}\cdot K}). \nonumber
\end{equation}
Then for any element $g \in G$ of the conformal group we have
\begin{equation}
G(gx_i) = \Big(\pi_1(dg_{x_1})\otimes\dots\otimes\pi_4(dg_{x_4})\Big) G(x_i). \nonumber
\end{equation}
Here $gx_i$ denotes the image of $x_i$ under the action of the global conformal transformation with $g$
and $dg_{x_i}$ was introduced in the paragraph preceding this theorem.}
\medskip

\noindent
{\sc Proof:}  It suffices to show that the claim holds for translations, rotations, dilations and the Weyl
inversion $w$, since these generate the whole conformal group. If $g$ is a translation the claim is clear
since $dg\equiv1$ and $G(x_i)$ is manifestly translation invariant. Assume now that $g$ is a dilation
$gx = e^\lambda x$. Then
\begin{align*}
G(e^\lambda x_i) & =\frac{1}{(e^\lambda x_{12})^{2\Delta_2} (e^\lambda x_{34})^{2\Delta_4}}\, \hat\rho(e^\lambda x_i)\,
F(e^{e^{-\lambda}Ix_{21}\cdot K}e^{e^\lambda x_{13}\cdot P}e^{e^{-\lambda}I x_{34}\cdot K})=\\[2mm]
& = e^{-2\lambda(\Delta_2+\Delta_4)}\frac{1}{x_{12}^{2\Delta_2} x_{34}^{2\Delta_4}}\, \hat\rho(x_i) \,
F(e^{\lambda D}e^{Ix_{21}\cdot K}e^{-\lambda D}e^{\lambda D}e^{x_{13}\cdot P}e^{-\lambda D}e^{\lambda D}
e^{I x_{34}\cdot K}e^{-\lambda D})=\\[2mm]
&=e^{-2\lambda(\Delta_2+\Delta_4)} e^{-\lambda(\Delta_1-\Delta_2+\Delta_3-\Delta_4)} G(x_i) =
\Big(\pi_1(dg_{x_1})\otimes\dots\otimes\pi_4(dg_{x_4})\Big) G(x_i). \nonumber
\end{align*}
Here we have used that $\hat \rho(e^\lambda x_i)  = \hat \rho(x_i)$. The $\hat \rho = \hat\rho_e$ was defined in eq.\ \eqref{eq:hatrho}.
In the second to last step we have employed the covariance properties \eqref{covariance} of the function
$F$ and the fact that $\pi_i'(D) = -\pi_i(D)$. Thereby we have established the claim in the case of
dilations.
\smallskip

Next, let $g=r$ be some global rotation of the $d$-dimensional space. In this case one finds that
\begin{align*}
G(rx_i) = \frac{1}{x_{12}^{2\Delta_2} x_{34}^{2\Delta_4}} \Big(1\otimes\pi_2(rs_{x_{12}}r^{-1}s_{e_d})
\otimes1\otimes\pi_4(rs_{x_{34}}r^{-1}s_{e_d})\Big) F(e^{r Ix_{21}\cdot K}e^{r x_{13}\cdot P}e^{rI x_{34}\cdot K})
\end{align*}
We used the fact that rotations $r \in G $ commute with the conformal inversion $I$. As in the case of
dilations, we can manipulate the argument of the function $F$ and insert its covariance properties to
obtain
\begin{align*}
& F(e^{r Ix_{21}\cdot K}e^{r x_{13}\cdot P}e^{rI x_{34}\cdot K}) =
F(re^{Ix_{21}\cdot K}r^{-1}re^{x_{13}\cdot P}r^{-1}re^{I x_{34}\cdot K}r^{-1})=\\[2mm]
& = \Big(\pi_1(r)\otimes \pi_2'(r)\otimes \pi_3(r)\otimes\pi_4'(r)\Big) F(e^{Ix_{21}
\cdot K}e^{x_{13}\cdot P}e^{I x_{34}\cdot K}).
\end{align*}
The factor that comes in front of $F$ is combined with the change in the term $\hat \rho$
\begin{equation}
\pi_2 (r s_{x_{12}}r^{-1} s_{e_d}) \pi_2'(r) = \pi_2(r s_{x_{12}}r^{-1} s_{e_d}s_{e_d} r s_{e_d})
= \pi_2(r) \pi_2(s_{x_{12}}s_{e_d}), \nonumber
\end{equation}
and similarly for $\pi_4$. Putting the last two calculations together, we conclude that
\begin{equation}
G(rx_i) = \Big(\pi_1(r)\otimes\dots\otimes\pi_4(r)\Big) G(x_i) =
\Big(\pi_1(dg_{x_1})\otimes\dots\otimes\pi_4(dg_{x_4})\Big) G(x_i). \nonumber
\end{equation}
This shows that the function $G$ obeys the same Ward identities for rotations.
\smallskip

It remains to understand special conformal transformations. Instead of addressing them directly,
we shall make use of the fact that translations and special conformal transformations are related
by the conformal inversion and prove the claim for the Weyl inversion $w$. In evaluating the
function $G$ at the points $gx_i = w x_i$ we shall exploit the following identity in the
conformal group
\begin{equation}\label{identity}
e^{y\cdot P} e^{-I y\cdot K} = w e^{wy\cdot P} s_{e_d} s_y |y|^{-2},
\end{equation}
which is derived in \cite{Dobrev:1977qv}, equations $(1.27a)-(1.27f)$. In the following
computation, $y$ denotes the point $wx$ that is obtained from $x$ by $w$. Note that this
differs from the standard conformal inversion with $I$ by an additional reflection of the
sign in the last component. With this in mind we find, using eq.\ \eqref{identity}
\begin{align*}
&  \hspace*{-5mm}
 F(e^{I y_{21}\cdot K}e^{y_{13}\cdot P}e^{I y_{34}\cdot K})=F(e^{I y_{21}\cdot K}
e^{I y_1\cdot K} e^{-Iy_1\cdot K}e^{y_{1}\cdot P}e^{-y_3\cdot P}e^{Iy_3\cdot K}
e^{-Iy_3\cdot K}e^{I y_{34}\cdot K})=\\[2mm]
& =F(e^{I y_{21}\cdot K}e^{I y_1\cdot K} |y_1|^2 s_{y_1} s_{e_d} e^{wy_1\cdot P}
wwe^{-wy_3\cdot P}s_{e_d}s_{y_3}|y_3|^{-2}e^{-Iy_3\cdot K}e^{I y_{34}\cdot K})=\\[2mm]
& =F(e^{I y_{21}\cdot K}e^{I y_1\cdot K} |y_1|^2 s_{y_1} s_{e_d} e^{x_{13}\cdot P}
s_{e_d}s_{y_3}|y_3|^{-2}e^{-Iy_3\cdot K}e^{I y_{34}\cdot K}).
\end{align*}
Our next step is to commute the dilation and rotation factors all the way to the left and
right position so that we can pull them out of the argument with the help of the covariance
law \eqref{covariance},
\begin{align*}
&  \hspace*{-5mm}
 F(e^{I y_{21}\cdot K}e^{y_{13}\cdot P}e^{I y_{34}\cdot K})
= x_1^{2\Delta_{12}} x_3^{2\Delta_{34}} F(e^{|y_1|^2 I y_{21}\cdot K}e^{y_1\cdot K}
s_{y_1} s_{e_d} e^{x_{13}\cdot P}s_{e_d}s_{y_3}e^{-y_3\cdot K}e^{|y_3|^2I y_{34}\cdot K})=\\[2mm]
& = x_1^{2\Delta_{12}} x_3^{2\Delta_{34}} F(s_{y_1} s_{e_d}
e^{(s_{e_d}s_{y_1}|y_1|^2 I y_{21}-Ix_1)\cdot K} e^{x_{13}\cdot P} e^{(Ix_3+s_{e_d}s_{y_3}
|y_3|^2I y_{34})\cdot K}s_{e_d}s_{y_3})=\\[2mm]
& = x_1^{2\Delta_{12}} x_3^{2\Delta_{34}} \Big(\pi_{12}(s_{y_1} s_{e_d})\otimes\pi_{34}(s_{y_3}
s_{e_d})\Big) F( e^{Ix_{21}\cdot K} e^{x_{13}\cdot P} e^{I x_{34}\cdot K})\ .
\end{align*}
In the final step we have used that
\begin{equation}
s_{e_d} s_{y_1} y_1^2 I y_{21} - I x_1 = I x_{21}, \nonumber
\end{equation}
and a similar equation in order to simplify the argument of the third exponential. Both
identities can be verified by a direct calculation. Notice further that
\begin{equation}
\frac{1}{y_{12}^{2\Delta_2} y_{34}^{2\Delta_4}} = \Big(\frac{x_1^2 x_2^2}
{x_{12}^2}\Big)^{\Delta_2} \Big(\frac{x_3^2 x_4^2}{x_{34}^2}\Big)^{\Delta_4}. \nonumber
\end{equation}
Putting all these pieces together, we have therefore shown that
\begin{align*}
& \hspace*{-5mm} G(y_i) =  \frac{1}{y_{12}^{2\Delta_2} y_{34}^{2\Delta_4}}
\Big(1\otimes\pi_2(s_{y_{12}}s_{e_d})\otimes1\otimes\pi_4(s_{y_{34}}s_{e_d})\Big)
F(e^{I y_{21}\cdot K}e^{y_{13}\cdot P}e^{I y_{34}\cdot K})=\\[2mm]
&=\frac{x_1^{2\Delta_1}x_2^{2\Delta_2}x_3^{2\Delta_3}x_4^{\Delta_4}}{x_{12}^{2\Delta_2}
x_{34}^{2\Delta_4}} \Big(\pi_1(s_{y_1}s_{e_d})\otimes\pi_2(s_{y_{12}}s_{y_1})\otimes
\pi_3(s_{y_3}s_{e_d})\otimes\pi_4(s_{y_{34}}s_{y_3})\Big) F\ .
\end{align*}
Whenever we write $F$ without an argument it is understood that is just the element $g(x_i)$.
To finish the proof, we make use of the following simple identity for reflections
\begin{equation}
s_{e_d} s_{y_{12}} s_{e_d} = s_{Ix_1 - Ix_2} = s_{x_1} s_{x_{12}} s_{x_2}. \nonumber
\end{equation}
Along with the property $s_{x}s_{e_d} = s_{e_d} s_y$, we use it to show
\begin{equation}
s_{y_{12}} = s_{e_d} s_{x_1} s_{x_{12}} s_{x_2} s_{e_d} = s_{y_1}
s_{e_d} s_{x_{12}} s_{e_{d}} s_{y_2}. \nonumber
\end{equation}
This allows us to evaluate $G(y_i)$ further
\begin{align*}
&\hspace*{-5mm} G(y_i) = \frac{x_1^{2\Delta_1}x_2^{2\Delta_2}x_3^{2\Delta_3}
x_4^{\Delta_4}}{x_{12}^{2\Delta_2}x_{34}^{2\Delta_4}} \Big(\pi_1(s_{e_d}s_{x_1})
\otimes\pi_2(s_{e_d}s_{x_2}s_{x_{12}}s_{e_d})\otimes\pi_3(s_{e_d}s_{x_3})\otimes
\pi_4(s_{e_d}s_{x_4}s_{x_{34}}s_{e_d})\Big) F=\\[2mm]
&=  x_1^{2\Delta_1}\dots x_4^{2\Delta_4}\pi_1(s_{e_d}s_{x_1})\otimes\dots\otimes
\pi_4(s_{e_d}s_{x_4}) G(x_i) = \Big( \pi_1(dg_{x_1}) \otimes\dots\otimes
\pi_4(dg_{x_4})\Big) G(x_i).
\end{align*}
Once again, we have arrived at the transformation rule under conformal inversions that
is obeyed by four-point functions of spinning fields in a conformal field theory,
see eq.\ \eqref{eq:Phiinv}. This completes the proof of the theorem.

\section{Correlators from Harmonic Analysis}

In Section 2 and the previous appendix we have seen how any $K \times K$ covariant function
on the conformal group gives a solution to the Ward identities. Now we will show that the
converse is also true, i.e.\ that every solution of the Ward identities arises in this way.
The discussion will closely follow that of \cite{Schomerus:2016epl}, only being a bit more explicit.

Let $G_4:(\mathbb{S}^d)^4\xrightarrow{}W_P$ be one solution of the Ward identities. It extends
to a unique function $F_4:G^4\xrightarrow{}W_P$ with the following properties

\begin{equation}\label{4-fold}
F_{4}: G^4\xrightarrow{}W_P,\ F_{4}(p_i g_i) = \Big(\bigotimes_{i=i}^{4}\pi_i(p_i)\Big)
F_{4}(g_i),\ F(e^{x_i\cdot P}) = G(x_i).  \nonumber
\end{equation}

Here, $p$ denotes an element of the parabolic subgroup $P\subset G$ which is generated by
rotations, dilations and special conformal transformations. Representations $\pi_i$ of $K$
are extended to those of $P$ by acting trivially with the special conformal transformations.
The space of functions with covariance properties of $F_4$ is by definition the tensor
product of (possibly non-unitary\footnote{Indeed, as is well-known, the representations appearing in physical Euclidean CFT correlators are not unitary from the $\Spin(1,d+1)$ point of view, but are continuations of unitary positive energy representations of $\tilde \SO(2,d)$.}) principal series representations $\rho_i$ of $G$. Each
of the principal series is realised on the space of left-covariant vector-valued functions
on the group. For details on the principal series, see \cite{Dobrev:1977qv,Schomerus:2016epl,
Schomerus:2017eny}. The four-fold tensor product will be denoted by $\rho=\bigotimes_{i=1}^4
\rho_i$, and its carrier space by $V$.

One can alternatively realise a principal series representation on the space of right covariant
functions. If we do this for the second two representations, the element $F_4$ is mapped to the
function

\begin{equation}\label{4-fold}
F_{2,\hat 2}: G^4\xrightarrow{}W_P,\ F_{2,\hat 2}(g_1,g_2,g_3,g_4) = F_4 (g_1,g_2,g_3^{-1},g_4^{-1}). \nonumber
\end{equation}

By the theorem 9.5 of \cite{Dobrev:1977qv}, the tensor product of two principal series representations
is isomorphic to a representation induced from the subgroup $K$ of rotations and dilations. The
isomorphism takes a function $H_2: G^2\xrightarrow{}V_1\otimes V_2$ which belongs to the first
space to the function

\begin{equation}\label{Mack}
H_1: G\xrightarrow{}V_1\otimes V_2,\ H_1(g) = H_2(g,gw),
\end{equation}

where $w$ is the Weyl inversion. Here, it is assumed that representations are realised on spaces of
right-covariant functions. For the case of left-covariant functions, the isomorphism takes the form
$H_1(g) = H_2 (g,wg)$.

Making use of these isomorphisms, we see that the four-fold tensor product $\rho$ can be realised on
the space of functions
\begin{equation}
F_{1,\hat 1}: G^2 \xrightarrow{} W_P,\ F_{1,\hat 1}(k_l g_1, g_2 k_r) =
\Big(\pi_{12}(k_l)\otimes\pi_{34}(k_r^{-1})\Big) F_{1,\hat 1}(g_1,g_2). \nonumber
\end{equation}
The relation with between $F_{1,\hat 1}$ and $F_{2,\hat 2}$ is given by
\begin{equation}
F_{1,\hat 1} (g_1,g_2) = F_{2,\hat 2}(g_1,w g_1,g_2,g_2 w). \nonumber
\end{equation}
To arrive at the space of $K \times K$ covariant functions we need to trace the conditions on $G_4$
implied by the Ward identities through the maps above. For $F_4$ they read
\begin{equation}
F_4(g_1 g,g_2 g,g_3 g,g_4 g) = F_4(g_1,g_2,g_3,g_4). \nonumber
\end{equation}
Therefore $F_{1,\hat 1}$ obeys $F_{1,\hat 1}(g_1,g_2) = F_{1,\hat 1}(g_1 g,g^{-1}g_2)$. That is,
$F_{1,\hat 1}$ depends only on the product of its two arguments. Let
\begin{equation}
F:G\xrightarrow{}W_P,\ F(g) = F_{1,\hat 1}(g,e), \nonumber
\end{equation}
The map $G_4\mapsto F$ establishes an isomorphism of vector spaces $V\xrightarrow{}\Gamma^{\pi_i}_\sigma$,
where $V$ is the space of solutions to the Ward identities. This completes the proof of the claim.

We end this appendix with a couple of remarks. The inverse of the map that we just constructed is of course
given by (\ref{maine}). This follows from the Ward identities. One could also arrive at the answer directly
by inverting each of the isomorphisms from above. In this procedure, the only non-trivial input is the
inverse of the isomorphism (\ref{Mack}), which is explicitly written in \cite{Dobrev:1977qv} and is
essentially based on the identity (\ref{identity}). It is therefore not surprising that this identity
played the crucial role in the proof of the Ward identities. We shall not the write the direct derivation
here since it adds little to the current discussion, but an enthusiastic reader is invited to verify that
it reproduces formula (\ref{maine}).

\section{Derivation of eqs.\ \eqref{eq:uz} and \eqref{eq:lambda}}

In this appendix we shall compute the coordinates $u_1(x_i)$ and $u_2(x_i)$ as well
as $\lambda_l(x_i)$ and $\lambda_r(x_i)$ in the Cartan decomposition of the elements
$g(x_i)\in G$ that we defined in eq.\ \eqref{eq:CartanKPK}. The results were stated
in eqs.\ \eqref{eq:uz} and \eqref{eq:lambda}. Since we can work in any faithful
representation of $G=\text{Spin}(1,d+1)$, let us take the vector representation by
$(d+2)\times (d+2)$ matrices.
\medskip

\noindent
{\bf Theorem:} \textit{Take four points belonging to Euclidean configuration space on $\mathbb{R}^d$, i.e.
	$$\left(x_a\right)_{a=1}^4 \in \left(\mathbb{R}^d\right)^4
	\backslash \{x_a \neq x_b \,\, \forall \, x_a \neq x_b, \,\,\, a,b = 1, \dots , 4\}.$$ Then for
	$d\geq 2$, the following decomposition holds\footnote{For almost all such $g$, more precisely
		the singular subvariety is of codimension $d$.}:
	\begin{align}
	g(x_i)&\colonequals e^{Ix_{21} \cdot K} e^{x_{13} \cdot P} e^{Ix_{34} \cdot K} =
	e^{\lambda^{(l)}D} \cdot r_l \left(\psi^{(l)}\right) \cdot
	e^{\frac{u_1}{2}(A_+-iA_-)+\frac{u_2}{2}(A_++iA_-)}  \cdot
	r_r \left(\psi^{(r)}\right) \cdot e^{\lambda^{(r)}D},
	\end{align}
	where, in particular,\footnote{The map \eqref{eq:z-to-u}:
		$(u_1, u_2) \twoheadrightarrow (z_1, z_2)$ is double-covering, since we can choose
		two different branches of a square root (doing it consistently for both $u_i$,
		the overall number of choices is two).}
	\begin{align}\label{eq:z-to-u}
	& e^{u_i} = 1-\frac{2}{z_i}\left(1+\sqrt{1-z_i}\right), \quad i=1,2,
	\end{align}
	and
	\begin{align}\label{eq:z-to-lambda}
	e^{2\lambda^{(l)}} = \frac{x_{12}^2 x_{14}^2}{x_{24}^2}\frac{1}{\sqrt{(1-z_1)(1-z_2)}},
	\quad  e^{2\lambda^{(r)}} = \frac{ x_{14}^2}{x_{13}^2 x_{34}^2}\frac{1}{\sqrt{(1-z_1)(1-z_2)}}.
	\end{align}
	Here we switched to usual cross-ratio variables
	\begin{align}
	U = z_1 z_2 = \frac{x_{12}^2 x_{34}^2}{x_{13}^2 x_{24}^2} , \quad  V = (1-z_1)(1-z_2)
	= \frac{x_{14}^2 x_{23}^2}{x_{13}^2 x_{24}^2},
	\end{align}
	so that Euclidean kinematics in the configuration space translates to $z_1=z_2^*,
	\quad z_i \in \mathbb{C}$. This factorization is unique, up to Spin$(d-2)$ redundancy
	reducing the number of independent angles among $\psi^{(l)}, \psi^{(r)}$ to
	$d(d-1) - (d-2)(d-3)/2 =(d^2+3d-6)/2$.}
\medskip

\noindent
{\sc Proof:} Clearly, both sides are elements of Spin$(1,d+1)$, so we just need to compare
the entries of these matrices.
One can directly see that the upper left $2\times 2$ block of the left hand side is of the form
\begin{equation}
\Gamma = \frac{1}{2}\left(
\begin{array}{cc}
x_{13}^2+\frac{x_{23}^2}{x_{12}^2}+\frac{x_{14}^2}{x_{34}^2}+\frac{x_{24}^2}{x_{12}^2 x_{34}^2}  & \
-x_{13}^2+\frac{x_{23}^2}{x_{12}^2}-\frac{x_{14}^2}{x_{34}^2}+\frac{x_{24}^2}{x_{12}^2 x_{34}^2}  \\
x_{13}^2-\frac{x_{23}^2}{x_{12}^2}+\frac{x_{14}^2}{x_{34}^2}-\frac{x_{24}^2}{x_{12}^2 x_{34}^2}   & \
-x_{13}^2+\frac{x_{23}^2}{x_{12}^2}+\frac{x_{14}^2}{x_{34}^2}-\frac{x_{24}^2}{x_{12}^2 x_{34}^2} \\
\end{array}
\right),\nonumber
\end{equation}
whereas the corresponding block of the right hand side looks as
\begin{equation}
\tilde\Gamma = \left(
\begin{array}{cc}
\cosh u_{12}^{+} \cosh \lambda_l \cosh \lambda_r +
\cosh u_{12}^{-} \sinh \lambda_l \sinh \lambda_r  & \
\cosh u_{12}^{+} \cosh \lambda_l \sinh \lambda_r +
\cosh u_{12}^{-} \sinh \lambda_l \cosh \lambda_r \\
\cosh u_{12}^{+} \sinh \lambda_l \cosh \lambda_r +
\cosh u_{12}^{-} \cosh \lambda_l \sinh \lambda_r  & \
\cosh u_{12}^{+} \sinh \lambda_l \sinh \lambda_r +
\cosh u_{12}^{-} \cosh \lambda_l \cosh \lambda_r\\
\end{array}
\right),\nonumber
\end{equation}
with $u_{12}^{\pm} = (u_1 \pm u_2)/2$.
Noticing that
\begin{align}
U=\frac{4}{\left(\Gamma_{11}-\Gamma_{22}\right)^2-\left(\Gamma_{12}-\Gamma_{21}\right)^2}, \quad V=
\frac{\left(\Gamma_{11}+\Gamma_{22}\right)^2-\left(\Gamma_{12}+\Gamma_{21}\right)^2}
{\left(\Gamma_{11}-\Gamma_{22}\right)^2-\left(\Gamma_{12}-\Gamma_{21}\right)^2}
\end{align}
and comparing to the left hand side, we obtain eq.\ \eqref{eq:z-to-u}.
Noticing that
\begin{align}
&4\frac{\sqrt{V}}{U} \cosh\left(2\lambda_r\right)=\Gamma_{11}^2+\Gamma_{12}^2-\Gamma_{21}^2-\Gamma_{22}^2,\\
&4\frac{\sqrt{V}}{U} \cosh\left(2\lambda_l\right)=\Gamma_{11}^2-\Gamma_{12}^2+\Gamma_{21}^2-\Gamma_{22}^2\nonumber
\end{align}
and comparing to the left hand side, we obtain eq.\ \eqref{eq:z-to-lambda}.

Computing the $(d^2+3d-6)/2$ Euler angles for rotational matrices $r_l$ and $r_r$ requires to compare
further matrix elements beyond the upper left corner. It is more difficult to obtain closed formulas
for these angles. In Section 4, however, we compute the six Euler angles for the 3-dimensional case,
using the 4-dimensional representation of the conformal group rather than the fundamental one.
Extending this type of analysis for higher dimensional conformal groups is possible, but beyond the
scope of this work.

\section{Euler Angles and Quaternions}

To understand the group-theoretic origin of the map \eqref{map-p} a bit better, let us first recall
that for the 3-dimensional conformal group, the following chain of group isomorphisms holds
\begin{align}\label{eq:spin3d-iso}
\text{Spin}_{\mathbb{R}}(1,4) \simeq \left(\mathcal{C}\ell_{\mathbb{R}}^0\right)^{\times}(1,4)/
\mathbb{R}^+ \simeq \mathcal{C}\ell_{\mathbb{R}}^{\times}(1,3)/\mathbb{R}^+ \simeq
\mathbb{H}^{2\times 2}/\mathbb{R}^+ \equiv \mathrm{SL}\left(2, \mathbb{H}\right).
\end{align}
Here $\mathcal{C}\ell^{0}_{\mathbb{R}}$ denotes an even part of the corresponding Clifford algebra\footnote{We 
use a 'plus sign' convention in definition of the Clifford algebra, i.e. such that the ideal to mod out of tensor 
algebra $T(V)$ of the underlying vector space $V$ is generated by relation $v\cdot v = q(v)\,\text{1}$, with 
$v\in V$ and $q$ quadratic form on $V$.}
over the real numbers and $(\mathcal{C}\ell^0)^{\times}_{\mathbb{R}}$ denotes its subgroup of
invertible elements. The first isomorphism follows from the definition of the Spin group and is
a low-dimensional exception,\footnote{To render it isomorphism for Euclidean conformal groups
	of $d>4$ dimensions, an additional condition should be imposed of taking only those elements of
	the Clifford group whose corresponding inner automorphisms preserve the vector space of gamma
	matrices. For low-dimensional Spin groups this is automatic, see \cite{lawson1989spin}.} while the projection
on its right hand side instructs us to take only elements of unit norm. The second comes from a
standard isomorphism of Clifford algebras implemented by taking products of a chosen gamma
matrix with the rest of those as a basis for the underlying vector space on the right hand side.
The third isomorphism, where the $\mathbb{H}^{2\times 2}$ denotes the group of invertible
$2\times 2$ quaternionic matrices, is implemented via the usual recursive construction of
gamma matrices
\begin{align*}
\mathcal{C}\ell_{\mathbb{R}}(1, 3) \simeq \mathcal{C}\ell_{\mathbb{R}}(0, 2) \otimes
\mathcal{C}\ell_{\mathbb{R}}(1, 1) \simeq \mathbb{H} \otimes \mathbb{R}(2) \simeq \mathbb{H}(2),
\end{align*}
where all isomorphisms are those of algebras and $\mathbb{R}(n),\, \mathbb{H}(n)$ denote the
corresponding $n\times n$ matrix algebra over reals/quarternions. Indeed, in the realization
\eqref{eq:4dfactorization} that we chose in Section 4 one can directly see eq.\
\eqref{eq:spin3d-iso} after multiplying the off-diagonal $2\times 2$ blocks by $i$.

Let us use the standard matrix presentation for a quaternion, i.e an element of $\mathbb{H}
\simeq \mathcal{C}\ell_{\mathbb{R}}(0, 2)$:\footnote{This isomorphism is not natural. We use
	the convention that the underlying vector space of $\mathcal{C}\ell_{\mathbb{R}}(0, 2)$ is
	spanned by basis vectors which are images of $\mathbf{J}$ and $\mathbf{K}$.}
\begin{align*}
X=x_0+x_1 \mathbf{I} +x_2 \mathbf{J} +x_3 \mathbf{K}=
\begin{pmatrix}
x_0+ix_1 & x_2+ix_3 \\
-x_2+ix_3 & x_0-ix_1
\end{pmatrix},
\end{align*}
where the imaginary units $\mathbf{I}, \mathbf{J}, \mathbf{K}$ all square to -1, anticommute
and satisfy $\mathbf{I}\mathbf{J}=\mathbf{K}$, $\mathbf{J}\mathbf{K}=\mathbf{I}$,
$\mathbf{K}\mathbf{I}=\mathbf{J}$. In this language, the rotation matrices $L$ and $L'$ (as
well as $R$ an $R'$) are unit quaternions (versors) related by the conjugation with
$\mathbf{I}$, $L' = -\mathbf{I} L \mathbf{I} \equiv L^{\mathbf{I}}$. In our chosen
conventions this can be also viewed as the involution $t$ of Clifford algebra $\mathcal{C}
\ell_{\mathbb{R}}(0, 2)$ reverting the signs of odd elements. Since $R$ and $L$ have unit
norm, $L^{-1}=\overline{L}$ and $R^{-1}=\overline{R}$, where the bar denotes the conjugation
anti-involution of $\mathbb{H}$.

Quaternion multiplication from the left  and from the right is clearly an $\mathbb{R}$-linear
operation, so it can be implemented by acting with $4\times 4$ real matrices on vectors of
quaternion parameters.  Explicitly, these are well-known maps,
\begin{align}
& v: \mathbb{H} \to \mathbb{R}^4, \quad X \mapsto (x_0, x_1, x_2, x_3)^T \nonumber\\
& \iota_L: \mathbb{H} \to \mathbb{R}(4), \quad  A \mapsto
\begin{pmatrix}
a_0 & -a_1 & -a_2 & -a_3 \\
a_1 & a_0 & -a_3 & a_2 \\
a_2 & a_3 & a_0 & -a_1 \\
a_3 & -a_2 & a_1 & a_0
\end{pmatrix} \\
& \iota_R: \mathbb{H} \to \mathbb{R}(4), \quad B \mapsto
\begin{pmatrix}
b_0 & -b_1 & -b_2 & -b_3 \\
b_1 & b_0 & b_3 & -b_2 \\
b_2 & -b_3 & b_0 & b_1 \\
b_3 & b_2 & -b_1 & b_0
\end{pmatrix},
\nonumber
\end{align}
where $A, B, X \in \mathbb{H}$.
The properties of these maps \cite{klaus1997quaternionic} (Lemma 1.23) are in particular such that
\begin{itemize}
	\item $\iota_L(A_1 A_2)=\iota_L(A_1)\iota_L(A_2)$, \,\, $\iota_R(B_1 B_2)=\iota_R(B_2)\iota_R(B_1)$
	\item $\iota_L(A)\iota_R(B)=\iota_R(B)\iota_L(A)$
	\item $v(AXB)=\iota_L(A)\iota_R(B) v(X)$.
\end{itemize}
One can understand them more conceptually by embedding both $\mathbb{H}$ and $\mathbb{R}^4$
into $\mathcal{C}\ell_{\mathbb{R}}(4,0)$ \cite{shao2017basisfree}. Let us introduce the twisted operation
$\overline{\iota_R} (B)=\iota_R \circ \overline{B}$, such that $\overline{\iota_R}(B_1 B_2)=
\overline{\iota_R}(B_1)\overline{\iota_R}(B_2)$.
Now we see that the map $v$ turns the algebra of quaternions considered as a bimodule over
itself into a left $\mathbb{H}\otimes\mathbb{H}$ module
\begin{align*}
(A, B).\, v(X) = \iota_L(A)\overline{\iota_R}(B) v(X).
\end{align*}
Since we can do that for any of the four quarternions in  $\mathbb{H}^{2\times 2}$, this
takes care of finding a map $\tilde{p}$ that has a property $\tilde{p}(r_l a r_r)= L \otimes
(R')^{-1} \tilde{p}(g_a)$ columnwise, see also \cite{Janovska}.
Requiring that also $\tilde{p}(e^{\lambda_1 D}r_l a r_r e^{\lambda_2 D})=  \tilde{p}(r_l a r_r)
\cdot D_1$ chooses an arrangement of the four $4\times 1$ blocks next to each other and fixes
$p$ uniquely, as in the formula \eqref{map-p}.


\end{document}